\documentclass[lettersize,journal]{IEEEtran}

\usepackage{subfigure, amsmath, rotating}
\usepackage{array, multirow}

\usepackage{epstopdf}
\usepackage{mathrsfs}
\usepackage{caption}
\usepackage{amssymb}
\usepackage{subfiles}
\usepackage{mathtools}

\usepackage{tikz}
\usepackage{float}
\usepackage{cite}
\usepackage{enumitem}
 
\usepackage{amsbsy, amsfonts, graphics, graphicx}
\usepackage{algorithmic}
\usepackage{algorithm}
\usepackage{amsthm}

\DeclareFontFamily{U}{mathc}{}
\DeclareFontShape{U}{mathc}{m}{it}%
{<->s*[1.03] mathc10}{}
\DeclareMathAlphabet{\mathscr}{U}{mathc}{m}{it}

\newtheorem{assumption}{Assumption}
\newtheorem{theorem}{Theorem}
\newtheorem{definition}{Definition}
\newtheorem{problem}{Problem}
\newtheorem{remark}{Remark}
\newtheorem{lemma}{Lemma}

\author{IEEE Publication Technology,~\IEEEmembership{Staff,~IEEE,}
\thanks{This paper was produced by the IEEE Publication Technology Group. They are in Piscataway, NJ.}
\thanks{Manuscript received April 19, 2021; revised August 16, 2021.}}

\begin{document}
\title{A Control Barrier Function Composition Approach for Multi-Agent Systems in Marine Applications}
\author{Yujia Yang, Chris Manzie, and Ye Pu
\thanks{Y. Yang, C. Manzie, and Y. Pu are with the Department of Electrical and Electronic Engineering, University of Melbourne, Parkville VIC 3010, Australia {\tt\small {yujyang1}@student.unimelb.edu.au, \{manziec,ye.pu\}@unimelb.edu.au}}
\thanks{Y. Yang is supported by the Melbourne Research Scholarship provided by the University of Melbourne; Y. Pu acknowledges support from the Australian Research Council via the Discovery Early Career Researcher Awards (DE220101527).}
}


\maketitle

\begin{abstract}
The agents within a multi-agent system (MAS) operating in marine environments often need to utilize task payloads and avoid collisions in coordination, necessitating adherence to a set of relative-pose constraints, which may include field-of-view, line-of-sight, collision-avoidance, and range constraints.
A nominal controller designed for reference tracking may not guarantee the marine MAS stays safe w.r.t. these constraints.
To modify the nominal input as one that enforces safety, we introduce a framework to systematically encode the relative-pose constraints as nonsmooth control barrier functions (NCBFs) and combine them as a single NCBF using Boolean composition, which enables a simplified verification process compared to using the NCBFs individually.
While other relative-pose constraint functions have explicit derivatives, the challenging line-of-sight constraint is encoded with the minimum distance function between the line-of-sight set and other agents, whose derivative is not explicit. 
Hence, existing safe control design methods that consider composite NCBFs cannot be applied.
To address this challenge, we propose a novel quadratic program formulation based on the dual of the minimum distance problem and develop a new theory to ensure the resulting control input guarantees constraint satisfaction.
Lastly, we validate the effectiveness of our proposed framework on a simulated large-scale marine MAS and a real-world marine MAS comprising one Unmanned Surface Vehicle and two Unmanned Underwater Vehicles.
\end{abstract}
 

\section{Introduction} \label{sec intro}
\begin{figure}[t!]
 \centering
\includegraphics[width=0.85\hsize]{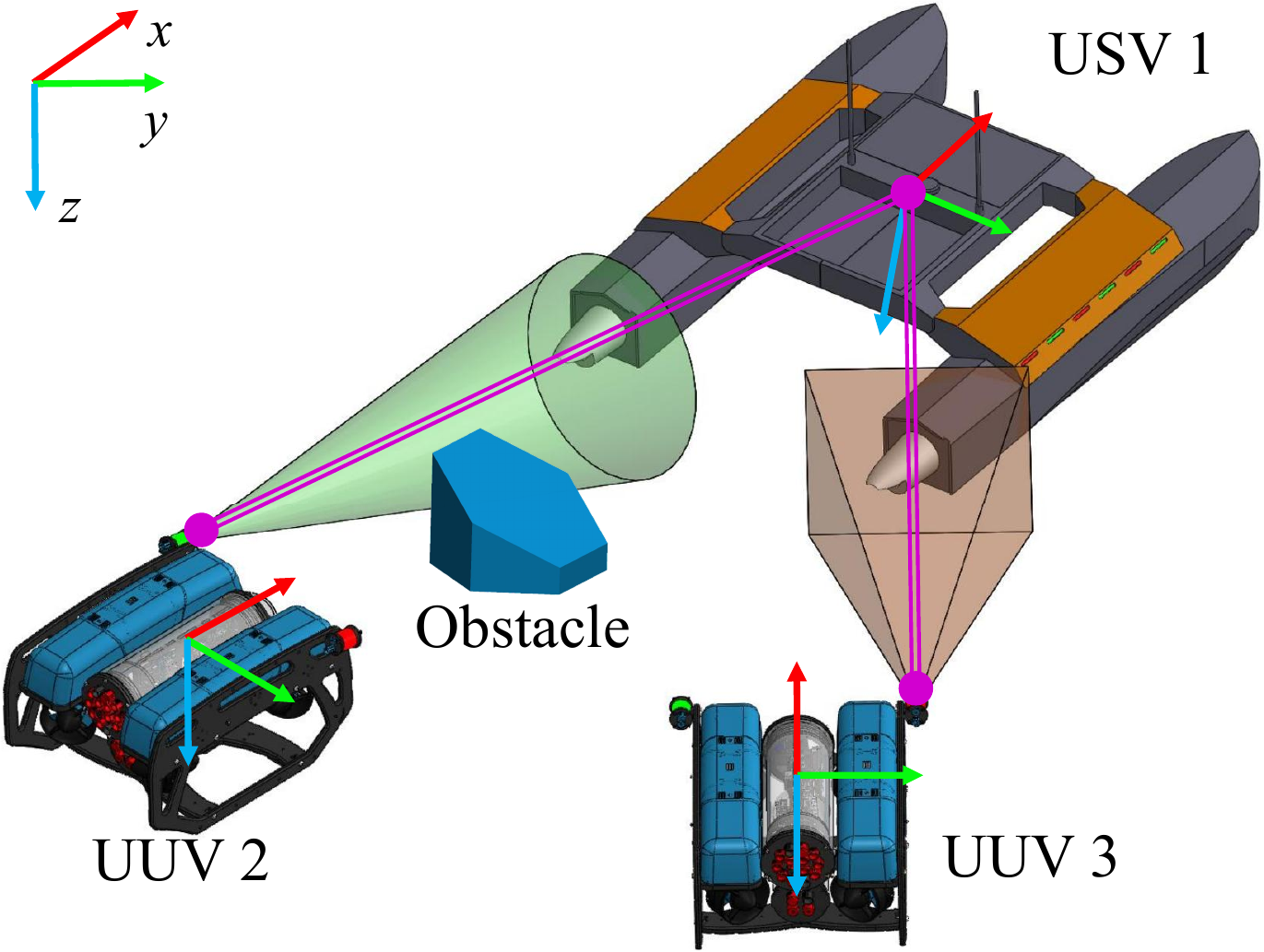}
\caption{A marine MAS with optical communication.}
\label{running example fig}
\end{figure}

A multi-agent system (MAS) can be advantageous in achieving maritime missions like ocean cartography and monitoring \cite{MMAS_app_survey, MMAS_app_5} compared to a group of single agents, for its robustness to single-point failure, ability to navigate in coordination, and utilize heterogeneous task payloads.
The functionality of common payloads like cameras, sonars, and communication devices can be characterized by a set of relative-pose constraints. For example, a camera must keep its target within its field-of-view (FOV) and range while ensuring an unobstructed line-of-sight (LOS). 
Furthermore, the requirement for coordination among agents complicates these constraints. Consider the marine MAS in Fig. \ref{running example fig}, where two unmanned underwater vehicles (UUVs) communicate with an unmanned surface vehicle (USV) through optical communication. Here, the USV and UUVs need to coordinate to fulfill FOV (green and red cones), LOS (purple double lines), and range constraints. Feasibility of the LOS constraint is potentially more challenging to guarantee as the LOS must remain unobstructed by other agents and the obstacle (blue polytope). Moreover, collision avoidance must be ensured.

Relative-pose constraints have been studied in broad contexts for their importance. 
%
%
The simplest form of relative-pose constraints considers just relative position. 
This covers inter-agent and agent-obstacle collisions and has been handled using approaches such as artificial potential fields \cite{MMAS_paper_1}.
An additional level of complexity is required to ensure FOV coverage, and to address constraints such as these, a controller based on dipolar reference vector fields was proposed \cite{MMAS_paper_3}.
Furthermore, formation control methods were proposed to impose relative-position constraints on MASs such that the agents maintain formations with rigid angles \cite{formation_control_1_angle_rigid} and rigid edge lengths \cite{formation_control_2_dist_rigid}.
One solution to ensure LOS connectivity for MAS requires mix-integer programs to guarantee collision avoidance between midpoints on the LOS set and obstacles \cite{opt_1_los_mix}.
In \cite{robot_1_vision}, FOV and LOS constraints were enforced for visual servoing robotic manipulators with safe velocity commands.
However, these methods do not consider the full range of constraints that are encountered in marine applications of MAS, as illustrated in Fig. 1, which include the relative-pose requirements to maintain communication between subsets of agents.

One class of methods for ensuring constraint satisfaction is the control barrier function (CBF) \cite{CBF_survey_ames}, which has historically been derived from smooth functions.
In \cite{cbf_3_wide_fov}, both FOV and collision avoidance constraints in a MAS were encoded as CBFs.
%
The authors of \cite{cbf_2_los_visual_servoing} ensured clear LOS for mobile visual sensors by forming a collision avoidance CBF between target points and obstacles in the image plane.
Since CBF unifies safety analysis under Nagumo's set-invariance theorem \cite{Nagumo1942berDL}, simultaneous consideration of multiple CBFs is possible.
In \cite{Xu2018ConstrainedCO, ShawCortez2022ARM}, conditions necessary for incorporating multiple constraint functions in safe control design through quadratic programs (QPs) were examined.
In \cite{Black2022AdaptationFV} multiple CBFs were composed into a single CBF for guaranteeing non-vanishing control authority.
Relative to other methods, incorporating a CBF into a controller provides a straightforward augmentation to encode and compose constraint functions.

Recently, the extension of CBF to nonsmooth constraint functions has attracted interest due to the potential for reduced conservatism in handling certain groups of constraints.
In \cite{magnus_lcs}, a composition approach was proposed to combine multiple barrier functions into a single nonsmooth barrier function.
This was later extended in \cite{magnus_ccta} to a nonsmooth control barrier function (NCBF) under the assumption of continuously differentiable component functions.
This is not directly applicable to the complete set of relative-pose constraints required for marine MAS applications.


The objective of this research is to develop a control framework that systematically enforces all relative-pose constraints in marine MASs, including the FOV, LOS, collision avoidance, and range constraints.
In the absence of a comprehensive method for enforcing these constraints, we propose to apply the Boolean composition introduced in \cite{magnus_lcs} (and extended by \cite{magnus_ccta}) to integrate the constraint functions into a composite NCBF.
Within these constraints, the LOS constraint function results from a minimum distance problem involving the LOS set and other agents and has no explicitly expressed derivatives that can be used to enforce constraint satisfaction.
%
%
To address these challenges, this paper makes the following contributions:
\begin{itemize}
    \item We encode the LOS constraint through a dual-based method and compose all the relative-pose constraints using Boolean composition. A novel QP is constructed for designing safe control inputs. The proposed composition is shown to guarantee constraint satisfaction.
    \item  Through simulation and experimentation on a marine MAS platform, the proposed framework is comprehensively validated.
\end{itemize}

\noindent \textbf{Notation:} 
For a set $\mathcal{S}$, let $\operatorname{co} \mathcal{S}$ denote its convex hull and $|\mathcal{S}|$ denote its cardinality.
For vectors $v, z \in \mathbb{R}^n$, let $\langle v, z\rangle$ be their dot product.
Let $v_{\{k\}}$ denote the $k$-th element of vector $v$.
Given a scalar $r\in \mathbb{R}_{+}$, the ball centered around $v$ is defined as $B(v,r) := \{ v' \in \mathbb{R}^n \mid \| v - v' \|^2 \leq r \}$.
Let $2^{\mathbb{R}^n}$ denote the power set of $\mathbb{R}^n$.
 

\section{Preliminaries}

This section reviews two NCBF methods that serve as the basis for our proposed control framework.
%

\subsection{Nonsmooth Analysis and NCBF}

We first provide some background on nonsmooth analysis and NCBFs.
Consider a control affine system 
\begin{align} \label{global dynamics}
    \dot{x} =f(x )+g(x ) u(x ), \; x \in \mathcal{D}, \;  u \in \mathcal{U},
\end{align}
where $f$ and $g$ are continuous, $\mathcal{D} \subset \mathbb{R}^n$, and $\mathcal{U} \subset \mathbb{R}^m$. 
When the control input $u(x)$ is discontinuous, the dynamics \eqref{global dynamics} becomes discontinuous too.
The Filippov operator can transform the discontinuous dynamics into a differential inclusion.

\begin{definition}[\cite{magnus_tac}]  \label{filipov definition}
The Filippov operator $K[f+$ $g u]: \mathbb{R}^n \rightarrow 2^{\mathbb{R}^n}$ w.r.t. \eqref{global dynamics} at $x$ is   
   \begin{align} \label{filipov}
  K[f+g u]\left(x\right)  := \operatorname{co} L[f+g u]\left(x\right),
   \end{align}
where $L: \mathbb{R}^n \rightarrow 2^{\mathbb{R}^n}$ is the map of limit points defined as
\begin{align}
 & L[f+g u]\left(x\right)   \\
 & =  \{\lim _{q \rightarrow \infty} f\left(x_q\right)+g\left(x_q\right) u\left(x_q\right) \mid x_q \rightarrow x, x_q \notin  S \cup \bar{S}_F \}, \nonumber
\end{align}
with $S$ being any set of Lebesgue measure zero in $\mathbb{R}^n$ and $\bar{S}_F$ being the zero-measure set where $f + g u$ is non-differentiable.
\end{definition}

Using the Filipov operator, we can define a Filippov solution $x(t)$ \cite{jorge_nonsmooth} which is an absolutely continuous function
$x:\left[0, T\right] \rightarrow \mathcal{D}$ that satisfies
\begin{align} \label{diff inclu definition}
    \dot{x}(t) \in F := K[f+g u]: \mathbb{R}^n \rightarrow 2^{\mathbb{R}^n}, \;  x(0)=x_0,
\end{align}
almost everywhere (a.e.) on $t \in\left[0, T\right]$, where $T$ is the time until which the solution $x(t)$ is defined. 
A function $f: \mathbb{R}^n \rightarrow \mathbb{R}^m$ is locally Lipschitz at $x$ if there exist $\delta, L>0$ such that $\left\|f\left(x_1\right)-f\left(x_2\right)\right\| \leq L\left\|x_1-x_2\right\|$ for all $x_1, x_2 \in B(x, \delta)$. Such functions can be used as candidate NCBFs, and their generalized gradients are defined as follows.

\begin{definition}[Definition 1, \cite{magnus_ccta}] \label{generalized gradient definition}
   Let $h: \mathbb{R}^n \rightarrow \mathbb{R}$ be Lipschitz continuous near $x$, and suppose $S$ is any set of Lebesgue measure zero in $\mathbb{R}^n$. Then, the generalized gradient $\partial h(x)$ is
    \begin{align}
        \partial h(x)=\operatorname{co}\{\lim _{i \rightarrow \infty} \nabla h\left(x_i\right) \mid x_i \rightarrow x, x_i \notin S \cup \bar{S}_h \},
    \end{align}
where $\bar{S}_h$ is the zero-measure set where $h$ is non-differentiable.
\end{definition}

\begin{definition}[Definition 2, \cite{magnus_ccta}] \label{candidate NCBF definition}
 A locally Lipschitz function $h: \mathcal{D} \rightarrow \mathbb{R}$, where $\mathcal{D}$ is an open, connected set, is a candidate NCBF if the safe set $\mathcal{C} :=\left\{x \in \mathcal{D} \mid h\left(x\right) \geq 0\right\}$ is nonempty.
\end{definition}

When NCBFs are included in the safe control design, the input $u(x)$ may become discontinuous. 
If a candidate NCBF is designed such that the resulting Filippov solution $x(t)$ satisfies
\begin{align} \label{valide ncbf condition}
    x(0) \in \mathcal{C} \Rightarrow x(t) \in \mathcal{C}, \;\; \textup{a.e. } t \in [0, T],
\end{align}
then, the candidate NCBF is also a valid NCBF.
A sufficient condition guaranteeing \eqref{valide ncbf condition} holds is:

\begin{theorem}[Theorem 3, \cite{magnus_lcs}] \label{bool main proposition}
Let $h: \mathcal{D} \rightarrow \mathbb{R}$ be locally Lipschitz function which is a candidate NCBF. Let $\Phi_f, \Phi_h: \mathcal{D} \subset \mathbb{R}^n \rightarrow$ $2^{\mathbb{R}^n}$ be set-valued maps such that
\begin{align} \label{prop 2 definition}
 F\left(x\right) \subset \operatorname{co} \Phi_f\left(x\right), \;\; \partial h\left(x\right) \subset \operatorname{co} \Phi_h\left(x\right),  
\end{align}
for all $x \in \mathcal{D}$. If there exists a locally Lipschitz extended class-$\mathcal{K}$ function ${\alpha}: \mathbb{R} \rightarrow \mathbb{R}$ such that for every $x \in \mathcal{D}$, $ z \in \Phi_h\left(x\right)$, and $v \in  \Phi_f\left(x\right)$,
\begin{align}  \label{prop 2 condition}
    \langle z, v\rangle \geq-{\alpha}\left(h\left( x \right)\right), 
\end{align}
then $h$ is a valid NCBF.
\end{theorem}

Next, we review two NCBF methods.

\subsection{Boolean Composition of Multiple Constraints via NCBFs} \label{bool method introduction}
 
The method in \cite{magnus_lcs}, extended by \cite{magnus_ccta}, allows the combination of multiple NCBFs into a single Boolean-NCBF (BNCBF) using Boolean operators defined below.

\begin{definition} \label{bool operators}
    For a pair of candidate NCBFs $h_1, h_2: \mathcal{D} \subset$ $\mathbb{R}^n \rightarrow \mathbb{R}$ and $x \in \mathcal{D}$, a candidate BNCBF is given by
\begin{align}
& h\left(x\right)=\min \left\{h_1\left(x\right), h_2\left(x\right)\right\}:=h_1 \wedge h_2  && (\textup{AND}), \label{bool op 1}\\
& h\left(x\right)=\max \left\{h_1\left(x\right), h_2\left(x\right)\right\}:=h_1 \vee h_2, && (\textup{OR}),  \label{bool op 2}\\
& h\left(x\right)=-h_1\left(x\right):=\neg h_1, &&  (\textup{NOT}).\label{bool op 3}
\end{align}
\end{definition}
By applying the above operators, multiple component functions $h_1, \cdots, h_k$, $k \geq 2$, can be composed as a single BNCBF,
\begin{align}
    h=\mathcal{B}\left(h_1, \ldots, h_k\right) 
\end{align}
where $\mathcal{B}$ denotes the nested Boolean operator. For an example, consider $h=\mathcal{B}\left(h_1,h_2,h_3, h_4\right) = (((h_1 \wedge h_2) \vee h_3) \vee \neg h_4)$, 
Since a BNCBF is also an NCBF, Theorem 1 may be applied for validating a candidate BNCBF.

\subsection{Collision Avoidance between Polytopic Agents via NCBF} \label{collision avoidance method introduction}

In \cite{dual_full}, the minimum distance function between polytopic agents was treated as a candidate NCBF for encoding a collision avoidance constraint.
%
%
Consider a pair of agents $\mathscr{a}$ and $\mathscr{b}$ whose geometries are represented as polytopes 
\begin{align}  \label{agent geometry}
    {\mathcal{P}}^i(x^i) := \{ p  \mid A^i(x^i)  p \leq b^i(x^i) \}, \; i \in \{\mathscr{a}, \mathscr{b}\},
\end{align}
where $p^i \in \mathbb{R}^{3}$ is position of the agent in 3-D Euclidean space. Let $\dot{x}^i =f^i(x^i ) + g^i(x^i ) u^i$, $i \in \{\mathscr{a}, \mathscr{b}\}$, represent their dynamics, respectively.

\begin{assumption}[Assumption 4 \cite{dual_full}]\label{polytope regularity assumptiosn}
For $i \in \{\mathscr{a}, \mathscr{b}\}$, it holds:
\begin{enumerate}[label=(\alph*)]
\item $A^i(x^i), b^i(x^i)$ are continuously differentiable $\forall x^i \in \mathcal{D}^i$.
\item $\forall x^i \in \mathcal{D}^i$, the set of active constraints at any vertex of $\mathcal{P}^i(x^i)$ are linearly independent.
\item $\mathcal{P}^i(x^i)$ is bounded, and thus compact, and has a nonempty interior for all $x^i \in \mathcal{D}^i$.
\end{enumerate}
\end{assumption}

The minimum distance function is defined as
\begin{subequations}\label{min distance of two polytopes}
\begin{align} 
&   h^{{\mathscr{a}}{\mathscr{b}}} (x^{\mathscr{a}},x^{\mathscr{b}})  :=  \min_{p, p'}    \| p - p' \| \\
  & \;\;\qquad\qquad \textup{ s.t. }  p  \in {\mathcal{P}}^{\mathscr{a}}(x^{\mathscr{a}}), \; p'  \in {\mathcal{P}}^{\mathscr{b}}(x^{\mathscr{b}}).
\end{align}
\end{subequations}
When Assumption \ref{polytope regularity assumptiosn} holds, $h^{\mathscr{a}\mathscr{b}} (x^{\mathscr{a}},x^{\mathscr{b}})$ is locally Lipschitz continuous and is a candidate NCBF.
Since the derivative $\dot{h}^{\mathscr{a}\mathscr{b}} (x^{\mathscr{a}},x^{\mathscr{b}})$ is not explicitly computable, Theorem 1 cannot be applied to verify the NCBF.
To address this issue, a dual-based formulation in \cite{dual_full} provided a computable lower bound of $\dot{h}^{\mathscr{a}\mathscr{b}} (x^{\mathscr{a}},x^{\mathscr{b}})$.
The dual problem of \eqref{min distance of two polytopes} is
\begin{subequations}  \label{dual collision avoidance problem}
 \begin{align}
   \underset{{\lambda}^{\mathscr{a}}, {\lambda}^{\mathscr{b}} }{\operatorname{max}} & \;   {L}^{\mathscr{a}\mathscr{b}} ({\lambda}^{\mathscr{a}}, {\lambda}^{\mathscr{b}} )  \\
   \text { s.t. } &  \; {\lambda}^{\mathscr{a}} A^{\mathscr{a}} (x^{\mathscr{a}} )+{\lambda}^{\mathscr{b}} A^{\mathscr{b}}  ({x}^{\mathscr{b}})=0,  {\lambda}^{\mathscr{a}}, {\lambda}^{\mathscr{b}} \geq 0,
\end{align}   
\end{subequations}
with ${L}^{\mathscr{a}\mathscr{b}}$ the Lagrangian function and $(\lambda^{\mathscr{a}*}, \lambda^{\mathscr{b}*})$ the optimal dual variables.
For given $(x^{\mathscr{a}}, x^{\mathscr{b}})$ and the corresponding $(\lambda^{\mathscr{a}*}, \lambda^{\mathscr{b}*})$ obtained from \eqref{dual collision avoidance problem}, the time derivative of ${L}^{\mathscr{a}\mathscr{b}}({\lambda}^{\mathscr{a}} ,{\lambda}^{\mathscr{b}} )$ denoted by $\dot{L}^{\mathscr{a}\mathscr{b}} ( u^{\mathscr{a}}, u^{\mathscr{b}},  \dot{\lambda}^{\mathscr{a}}, \dot{\lambda}^{\mathscr{b}} )$ is
\begin{align} \label{dual cost definition}
 & \dot{L}^{\mathscr{a}\mathscr{b}}(\cdot) =  -\frac{1}{2} {\lambda}^{\mathscr{a}} A^{\mathscr{a}} ({x}^{\mathscr{a}}) A^{\mathscr{a}} ({x}^{\mathscr{a}})^\top \dot{{\lambda}}^{p \top}  -\dot{\lambda}^{\mathscr{a}} b^{\mathscr{a}}({x}^{\mathscr{a}})-{\lambda}^{\mathscr{a}} \dot{b}^{\mathscr{a}}({x}^{\mathscr{a}}) \nonumber \\
 &  -\frac{1}{2} {\lambda}^{\mathscr{a}} A^{\mathscr{a}} ({x}^{\mathscr{a}}) \dot{A}^{\mathscr{a}}({x}^{\mathscr{a}}, u^{\mathscr{a}})^\top {\lambda}^{p \top}  -\dot{\lambda}^{\mathscr{b}} b^{\mathscr{b}}(x^{\mathscr{b}})-{\lambda}^{\mathscr{b}} \dot{b}^{\mathscr{b}}(x^{\mathscr{b}})
\end{align}
with $\dot{A}^i({x}^i, u^i) = {L}_{f^i}A^i({x}^i) + {L}_{g^i}A^i({x}^i) {u}^i$, and $(\dot{{\lambda}}^{\mathscr{a}}, \dot{{\lambda}}^{\mathscr{b}})$ the time derivatives of $( {{\lambda}}^{\mathscr{a}*}, {{\lambda}}^{\mathscr{b}*})$.
\begin{lemma}[Lemma 10 \cite{dual_full} (with reformed notations)]\label{lowerbound prop}
Let  
\begin{subequations} \label{dual collision avoidance derivative problem}
\begin{align} 
  g^{\mathscr{a}\mathscr{b}}&(x^{\mathscr{a}} ,x^{\mathscr{b}},  u^{\mathscr{a}},u^{\mathscr{b}})    := \max_{ \dot{{\lambda}}^{\mathscr{a}}, \dot{{\lambda}}^{\mathscr{b}} } \dot{L}^{\mathscr{a}\mathscr{b}} (u^{\mathscr{a}},u^{\mathscr{b}}, \dot{\lambda}^{\mathscr{a}}, \dot{\lambda}^{\mathscr{b}} )  \label{LP a} \\
& \textup { s.t. } \; \dot{\lambda}^{\mathscr{a}} A^{\mathscr{a}} (x^{\mathscr{a}} )+{\lambda}^{\mathscr{a}*} \dot{A}^{\mathscr{a}}({x}^{\mathscr{a}}, u^{\mathscr{a}} )   
 \nonumber \\
 & \quad\,  +\dot{\lambda}^{\mathscr{b}} A^{\mathscr{b}}(x^{\mathscr{b}})+{\lambda}^{\mathscr{b}*} \dot{A}^{\mathscr{b}}(x^{\mathscr{b}}, u^{\mathscr{b}})=0,  \label{LP b } \\
& \quad\, \dot{{\lambda}}^{\mathscr{a}}_{\{k^{\mathscr{a}}\}}\geq 0, \; \dot{{\lambda}}^{\mathscr{b}}_{\{k^{\mathscr{b}}\}}\geq 0 , (k^{\mathscr{a}}, k^{\mathscr{b}}) \in  K^{0}(x^{\mathscr{a}}, x^{\mathscr{b}}), \label{LP c}
\end{align}    
\end{subequations}
where $K^{0}(x^{\mathscr{a}}, x^{\mathscr{b}}) := \{ (k^{\mathscr{a}}, k^{\mathscr{b}}) \mid {\lambda}^{\mathscr{a}*}_{\{k^{\mathscr{a}}\}} = 0, {\lambda}^{\mathscr{b}*}_{\{k^{\mathscr{b}}\}} = 0\}$, and $\dot{{\lambda}}^{\mathscr{a}}$ and $\dot{{\lambda}}^{\mathscr{b}}$ are derivatives of the optimal dual variables ${{\lambda}}^{l*}$ and ${{\lambda}}^{o*}$.
If Assumption \ref{polytope regularity assumptiosn} holds, then, for a.e. on $t \in [0,T]$,
\begin{align} \label{lowerbound equation}
    \dot{h}^{{\mathscr{a}}{\mathscr{b}}} (x^{\mathscr{a}}(t),x^{\mathscr{b}}(t)) \geq g^{\mathscr{a}\mathscr{b}}(x^{\mathscr{a}}(t) ,x^{\mathscr{b}}(t),  u^{\mathscr{a}}(t),u^{\mathscr{b}}(t)).
\end{align}
\end{lemma}

\section{Problem Statement} 

In this work, we consider control of a marine MAS consisting of $N$ agents, each with control-affine dynamics
\begin{align} \label{individual agent dynamics}
  \dot{x}^i =f^i(x^i )+g^i(x^i ) u^i , \;   i \in \mathcal{N} =\{1, \ldots, N\} ,
\end{align}
where $x^i \in \mathcal{D}^i \subset \mathbb{R}^n$ is the state, which contain pose (position $p^i$ and orientations) information of the agent.
The functions $f^i: \mathcal{D}^i \rightarrow \mathbb{R}^n$ and $g^i: \mathcal{D}^i \rightarrow \mathbb{R}^{n \times m}$ are continuous, $u^i  \in \mathcal{U}^i \subset \mathbb{R}^m$, and the geometry of agent $i$ is ${\mathcal{P}}^i(x^i)$ defined in \eqref{agent geometry} and satisfies Assumption \ref{polytope regularity assumptiosn}.
The marine MAS dynamics can be written as the system in \eqref{global dynamics} with $x = [{x^1}^{\top}, \cdots, {x^N}^{\top}]^{\top}$, $u = [u^{\top}_1, \cdots, u^{\top}_N]^{\top}$, $\mathcal{D} := \mathcal{D}^1 \times \cdots \times \mathcal{D}^N$, and $\mathcal{U} := \mathcal{U}^1 \times \cdots \times \mathcal{U}^N$.

\begin{assumption} \label{D and U assumptions}
    $\mathcal{D}$ is open, connected; $\mathcal{U}$ is convex.
\end{assumption}

The assumption on $\mathcal{D}$ is necessary for constructing candidate NCBFs, cf. Definition \ref{candidate NCBF definition}, while the assumption on $\mathcal{U}$ is common for marine MASs.
Next, we introduce a set of relative-pose constraints that emerge in marine MAS and summarize the control objective we consider.

\subsection{Relative-Pose Constraints in marine MAS} \label{problem definitions} 

For a MAS, like in Fig. \ref{running example fig}, to maintain a communication network a range constraint can enforce proximity between agents. Similarly, a  LOS constraint can guarantee an unobstructed communication channel between agents \cite{MMAS_survey_2}, and a FOV constraint can ensure an agent equipped with a forward-looking sonar maintains a clear view of a target.
Each of these constraints is formulated below, but all need to be simultaneously enforced.
 
\vspace{6pt}

\noindent \textbf{Field-of-View Constraint:} This constraint enforces agent $j$ to remain within the FOV of agent $i$ and is defined as
\begin{align} \label{FOV constraint definition}
  h^{ij}_{\textup{fov}}(x^i,x^j) := -\|A^i_{\textup{fov}} p^{ij} +b^i_{\textup{fov}}\| + {c^{i \top}_{\textup{fov}}} p^{ij} +d^i_{\textup{fov}}  \geq 0
\end{align}
where $p^{ij} = {R}\left( x^i \right)(p^j - p^i)$ is the vector pointing from agent $i$ to agent $j$, with ${R}\left( x^i \right)$ being the rotation matrix from the world frame to that of agent $i$.  
Constraint \eqref{FOV constraint definition} is a second-order-cone constraint in $p^{ij}$, where $A^i_{\textup{fov}} \in \mathbb{R}^{3 \times 3}$, $b^i_{\textup{fov}}, c^i_{\textup{fov}} \in \mathbb{R}^{3}$, and $d^i_{\textup{fov}} \in \mathbb{R}$ can be chosen to express a wide class of cone-shaped FOVs, including the ellipsoidal (green) and polyhedral (red) cones in Fig. \ref{running example fig}.

\vspace{6pt}

\noindent \textbf{Range Constraint:} The range constraint enforces the relative distance between agent $i$ and $j$ to remain in a range $[\underline{r}^{i}_{\textup{rng}}, \overline{r}^{i}_{\textup{rng}}]$, $\overline{r}^{i}_{\textup{rng}} \geq \underline{r}^{i}_{\textup{rng}} > 0$, and can be decomposed into two constraints:
\begin{align}
  &  \underline{h}^{ij}_{\textup{rng}}(x^i,x^j) :=  \| {p}^j - p^i \| -  \underline{r}^{i}_{\textup{rng}} \geq 0, \label{min range constraint definition} \\
  &  \overline{h}^{ij}_{\textup{rng}}(x^i,x^j) :=  \overline{r}^{i}_{\textup{rng}} - \| {p}^j - p^i \|  \geq 0. \label{max range constraint definition}
\end{align}

\noindent \textbf{Collision Avoidance Constraint:} For collision avoidance between agent $i$ and $j$, we define their minimum distance function as
\begin{align} \label{collision avoidance min dist definition}
  h^{ij }_{\textup{ca}}  (x^i,x^j ) &  :=  \min_{p,p'} \| p - p' \| -  r^{ij}_{\textup{ca}} \\
  & \textup{s.t.} \; p \in \mathcal{P}^{i}(x^i), \; p' \in \mathcal{P}^j(x^j) \nonumber
\end{align}
which is offset by a safe distance $r^{ij}_{\textup{ca}} > 0$.
We then require 
\begin{align} \label{collision avoidance constraint definition}
    h^{ij}_{\textup{ca}}  (x^i,x^j)  \geq 0.
\end{align}

\noindent \textbf{Line-of-Sight Constraint:} 
Let the LOS set between agent $i$ and $j$ be defined as 
\begin{align} \label{LOS set definition}
 &   \mathcal{P}^{ij}_{\textup{los}}(x^i,x^j) := \{ {p}  \mid  {p} = \alpha {p}^j + (1-\alpha) p^i, \alpha \in [0,1]  \},
\end{align}
and let an additional agent (or obstacle) be denoted as agent $k$ with geometry $\mathcal{P}^k(x^k)$.
Our goal is to ensure the LOS is not occluded by agent $k$, i.e.
\begin{align} \label{LOS obstacle set definition}
  &  \mathcal{P}^{ij}_{\textup{los}}(x^i,x^j) \cap \mathcal{P}^k(x^k) = \varnothing.
\end{align}
To encode the above requirement as a constraint function, we define the minimum distance between $\mathcal{P}^{ij}_{\textup{los}}(x^i,x^j)$ and $\mathcal{P}^k(x^k)$, offset by a safe distance $r^{ijk}_{\textup{los}} \geq 0$, as
\begin{align} \label{LOS min dist definition}
  h^{ijk}_{\textup{los}}  (x^i,x^j,x^k ) &  :=  \min_{p,p'} \| p - p' \| -  r^{ijk}_{\textup{los}} \\
  & \textup{s.t.} \; p \in \mathcal{P}^{ij}_{\textup{los}}(x^i,x^j), \; p' \in \mathcal{P}^k(x^k) \nonumber
\end{align}
and require 
\begin{align} \label{LOS constraint definition}
    h^{ijk}_{\textup{los}}  (x^i,x^j,x^k )  \geq 0.
\end{align}

Since $h^{ijk}_{\textup{los}}$ is the result of the minimization problem \eqref{LOS min dist definition}, its time-derivative cannot be explicitly expressed as $\dot{h}_i\left( x \right) := \mathcal{L}_{f}h_i(x) + \mathcal{L}_{g} h_i(x) u(x)$, where $\mathcal{L}_{(\star)}(\cdot)$ represents the Lie derivative of $(\cdot)$ along $(\star)$ \cite{CBF_survey_ames}. A CBF design that allows $h^{ijk}_{\textup{los}}$ to be enforced is necessary.

To address the challenge of encoding \eqref{LOS constraint definition}, we use the NCBF method introduced in Section \ref{collision avoidance method introduction}.
However, the LOS set ${\mathcal{P}}^{ij} (x^i,x^j)$ is a line segment with an empty interior and does not satisfy Assumption \ref{polytope regularity assumptiosn}.
To address this issue, we introduce the following non-restrictive assumption.

\begin{assumption} \label{over bound assumption}
There exists a polytope $\overline{\mathcal{P}}^{ij} (x^i,x^j)$ that satisfies Assumption \ref{polytope regularity assumptiosn} and $\mathcal{P}^{ij}_{\textup{los}}(x^i, x^j) \subseteq  \overline{\mathcal{P}}^{ij} (x^i,x^j)$.
\end{assumption}

\noindent We define a new minimum distance function that is amenable to the method in Section \ref{collision avoidance method introduction}, with $\overline{\mathcal{P}}^{ij} (x^i,x^j)$: 
\begin{align} \label{new LOS min dist definition}
  \overline{h}^{ijk}_{\textup{los}}  (x^i,x^j,x^k ) &  :=  \min_{p,p'} \| p - p' \| -  r^{ijk}_{\textup{los}} \\
  & \textup{s.t.} \; p \in \overline{\mathcal{P}}^{ij}_{\textup{los}}(x^i,x^j), \; p' \in \mathcal{P}^k(x^k). \nonumber
\end{align}

\begin{lemma}
Suppose Assumption \ref{over bound assumption} holds.
If $\overline{h}^{ijk}_{\textup{los}}  (x^i,x^j,x^k) \geq 0$, then $h^{ijk}_{\textup{los}}  (x^i,x^j,x^k) \geq 0$.

\begin{proof}
    The proof follows from the fact that $\overline{\mathcal{P}}^{ij} (x^i,x^j) \cap {\mathcal{P}}^k (x^k)  = \varnothing   \Rightarrow     {\mathcal{P}}^{ij} (x^i,x^j) \cap \mathcal{P}^{k}(x^k)  = \varnothing$.
\end{proof}
\end{lemma}

In the above, the FOV and range constraints can be encoded as smooth CBFs, while the CA and LOS constraints can be encoded through the NCBF method in Section \ref{collision avoidance method introduction}.
Now let $I_s$ and $I_n$ respectively be the total number of smooth CBFs and NCBFs applicable to marine MAS applications.
Let $\mathcal{I}_s := \{1,\cdots, I_s\}$ and $\mathcal{I}_n := \{I_s + 1,\cdots, I_s + I_n\}$.
We now introduce the ordered index set $\mathcal{I} := \mathcal{I}_{s} \cup \mathcal{I}_{n}$ which contains all the grouped constraint indices.

\subsection{Control Objective} \label{multiple cbf challenges}

In addition to satisfying the relative-pose constraints, a marine MAS needs to achieve navigation tasks, such as reaching reference states $x_r \in \mathcal{D}$. 
Suppose a nominal controller $u_{r}(x): \mathcal{D} \rightarrow \mathcal{U}$ satisfying the following condition is provided:
Let all $x_r \in \mathcal{D}$ be steady states of \eqref{global dynamics} with $u(x) = u_r(x)$. There exist a class-$\mathcal{K} \mathcal{L}$ function $\beta$ and a class-$\mathcal{K}$ function $\gamma$ such that for any $x\left(t_0\right) \in \mathcal{D}$, the solution $x(t)$ exists for all $t \geq t_0$ and satisfies
\begin{align} \label{ref input condition}
  \|x(t) - x_r\| \leq & \beta\left(\left\|x\left(t_0\right) - x_r \right\|, t-t_0\right)  \nonumber \\
   &  +\gamma(\sup _{t_0 \leq \iota \leq t}\|u_r(x(\iota))\|).
\end{align}
 Note that this condition can be satisfied using existing navigation controllers such as described in \cite{Zhang2023AdaptiveNC, Wang2023ModelPT}.

We aim to develop a systematic framework for enforcing the constraints $h_i(x(t)) \geq 0, \; \forall i \in \mathcal{I}$ while leveraging $u_{r}(x)$ to achieve navigation requirements.
Let $\mathcal{C}_i := \{ x\in \mathcal{D} | h_i(x) \geq 0\}$ be the safe set corresponding to the $i^{\textup{th}}$ constraint.
Then, if there exists extended class-$\mathcal{K}$ functions $\alpha_{i}$, $i \in \mathcal{I}$, such that the CBF-QP
\begin{subequations}  \label{multi-CBF-QP}
    \begin{align}  
& u(x)=\underset{u \in\mathcal{U}}{\operatorname{argmin}}  \left\|u-u_{r}(x)\right\|^2 \\
& \qquad\qquad \text { s.t. } \dot{h}_i(x) \geq-\alpha_{i}\left(h_i(x)\right), \quad \forall i \in \mathcal{I}, 
\end{align}
\end{subequations}
is feasible for all $x \in \mathcal{D}$, the input $u(x(t))$ guarantees
\begin{align} \label{multi-CBF-QP conditions}
    x(0) \in \mathcal{C}_i \Rightarrow x(t) \in \mathcal{C}_i, \; i \in \mathcal{I}, \; \forall t \geq 0.
\end{align}
However, designing a set of $\alpha_{i}, i \in \mathcal{I}$ may be challenging since each element of the set needs to be non-contradicting so that \eqref{multi-CBF-QP} is feasible.
%
Composition methods \cite{Black2022AdaptationFV}, \cite{magnus_lcs}, \cite{ames_composition_paper} that compose multiple CBFs as a single CBF can simplify the design process. However, these existing methods do not address the case where NCBFs are involved in the composition.
Consequently, the overall objective is captured with the following problem statement.

\begin{problem} \label{problem 1}
Suppose Assumptions \ref{polytope regularity assumptiosn} and \ref{D and U assumptions} hold.
Design a composite CBF $h_g$ such that the corresponding safe set $\mathcal{C}_g := \{ x \in \mathcal{D} \mid h_g \geq 0 \}$ satisfies $\mathcal{C}_g  \subset \mathcal{C}_1 \times \cdots \times \mathcal{C}_I$, and develop a method to obtain a safe control input $u(x(t))$ that minimally modifies $u_r(x(t))$ and ensures the system \eqref{global dynamics} satisfies
\begin{align} \label{problem 1 main condition}
    x(0) \in \mathcal{C}_g \Rightarrow x(t) \in \mathcal{C}_g, \; \forall t \geq 0.
\end{align}
\end{problem}

\section{Main Results}\label{main results}

Problem 1 requires the composition of multiple NCBFs.
In this section, we introduce a composition method to systematically handle the constraints $h_i, \; \forall i \in \mathcal{I}$.
We then utilize a QP-based safe control design to guarantee constraint satisfaction for the marine MAS.

\subsection{Composing Relative-Pose Constraints as BNCBF}
 
Using the $\wedge$ operations in \eqref{bool op 1}, we can compose $h_i(x)$, $i \in \mathcal{I}$ as the following BNCBF:
 \begin{align} \label{final bool definition}
    h_g := \mathcal{B}(h_1, \cdots, h_{{I}_{s}},  h_{{I}_{s}+1}, \cdots,  h_{{I}_{s} + {I}_{n}}). 
\end{align}
The corresponding safe set is
\begin{align} \label{global safe set}
    \mathcal{C}_g = \{ x \in \mathcal{D}  \mid h_g(x) \geq 0 \},
\end{align}
which contains all states that satisfy the relative-pose constraints encoded by $h_g$ and satisfies $\mathcal{C}_g  = \mathcal{C}_1 \times \cdots \times \mathcal{C}_I$.

\begin{assumption} \label{non empty assumption}
    $\mathcal{C}_g$ is nonempty.
\end{assumption}

\begin{assumption} \label{diff asumption}
     $\partial h_i(x) = \nabla h_i(x)$, $i \in \mathcal{I}_{s}$, for all $x \in \mathcal{C}_g$.
\end{assumption}
The derivatives of $h^{ij}_{\textup{fov}}$, $\underline{h}^{ij}_{\textup{rng}}$, and $\overline{h}^{ij}_{\textup{rng}}$ are undefined only when $\| p^i - p^j \| = 0$, which is prevented if $h^{ij}_{\textup{ca}} \geq 0$. 
Thus, Assumption \ref{diff asumption} holds for all $x \in \mathcal{C}_g$ if Assumption \ref{non empty assumption} holds.

Theorem \ref{bool main proposition} provides a sufficient condition \eqref{prop 2 condition} for verifying $h_g$ as a valid NCBF.
When $\mathcal{I}_{n} = \varnothing$, \cite{magnus_ccta} constructs a QP-based safe control design that guarantees the sufficient condition \eqref{prop 2 condition} holds, provided that the QP is feasible over $\mathcal{C}_g$.
A key condition underlying the QP is $\Phi_h\left(x\right)$ in \eqref{prop 2 definition} can be constructed solely from computable derivatives $\nabla h_i$, $i \in \mathcal{I}_{s}$.
When $\mathcal{I}_{s} = \varnothing$ and $\mathcal{I}_{n}$ is singleton (contains only one function), \cite{dual_full} constructs a QP using the lower bound proposed in Lemma \ref{lowerbound prop} for validating $h_g$.
To encode the relative-pose constraints in Problem \ref{problem 1}, we consider the more general case where both $\mathcal{I}_{s}$ and $\mathcal{I}_{n}$ are nonempty and non-singleton. This case is difficult to address because $\Phi_h\left(x\right)$ includes $\partial h_j$, $j \in \mathcal{I}_{n}$ as well, which are set-valued maps that may not have explicit expressions. 
We address this challenge by constructing a QP-based safe control design method.

\subsection{Safe Control Design through BNCBF-QP}

To guarantee $h_g$ as a valid NCBF, we construct a BNCBF-QP for safe control design and prove the resulting controller satisfies the sufficient condition \eqref{prop 2 condition} in Proposition \ref{bool main proposition}.
Using a similar approach to \cite{magnus_ccta}, a BNCBF-QP is defined as:
\begin{subequations}   \label{BNCBF QP}
\begin{align} 
 & {u}^*({x}) \in \underset{ {u}\in \mathcal{U}, \dot{{\lambda}}^{\mathscr{a}}_{j}, \dot{{\lambda}}^{\mathscr{b}}_{j} }{\operatorname{argmin}}\left\|{u}-{u}_{r}\right\|_Q  \label{qp cost} \\
  &  \text {s.t. }\langle \nabla h_i(x) , f(x) +g(x)u \rangle \geq -\alpha({h}_g(x) ) ,      \label{diff cons a}  \allowdisplaybreaks \\
&  \quad\quad\,  \dot{L}_j (u, \dot{{\lambda}}^{\mathscr{a}}_{j}, \dot{{\lambda}}^{\mathscr{b}}_{j} ) \geq -\alpha({h}_g(x)) ,   \label{poly cons a}  \allowdisplaybreaks \\
& \quad\quad\,  \dot{{\lambda}}^{\mathscr{a}}_{j} A^{\mathscr{a}}_{j}({x}  )+{\lambda}_{j}^{\mathscr{a}*} \dot{A}^{\mathscr{a}}_{j}(x,u) \nonumber \allowdisplaybreaks \\
& \quad\quad\, +\dot{{\lambda}}^{\mathscr{b}}_{j} A^{\mathscr{b}}_{j}({x})+{\lambda}^{\mathscr{b}*}_{j} \dot{A}^{\mathscr{b}}_{j}(x,u)=0, \label{poly cons b} \\
& \quad\quad\, \dot{{\lambda}}^{\mathscr{a}}_{j,\{k^{\mathscr{a}}\}}\geq 0, \; \dot{{\lambda}}^{\mathscr{b}}_{j,\{k^{\mathscr{b}}\}}\geq 0, \;\;  (k^{\mathscr{a}}, k^{\mathscr{b}}) \in K^{\epsilon_2}_j(x), \label{poly cons c} \\
&  \quad\quad\, i \in \mathcal{I}^{\epsilon_1}_{s}(x), \;\; j \in \mathcal{I}^{\epsilon_1}_{n}(x). \nonumber
\end{align}    
\end{subequations}
Inspired by \cite{magnus_ccta}, we introduce almost-active index sets $\mathcal{I}^{\epsilon_1}_{s}(x)$ and $\mathcal{I}^{\epsilon_1}_{n}(x)$, defined as
\begin{align}
   & \mathcal{I}^{\epsilon_1}_{s}\left(x\right) :=\left\{ i \in \mathcal{I}_s \mid \left|h_i \left(x\right)-h_g\left(x\right)\right| \leq \epsilon_1\right\}, \\
    &  \mathcal{I}^{\epsilon_1}_{n}\left(x\right) :=\left\{ j \in \mathcal{I}_n  \mid \left|h_j \left(x\right)-h_g\left(x\right)\right| \leq \epsilon_1\right\},
\end{align}
with $\epsilon_1 > 0$ a small constant, to account for the nonsmoothness in $h_g$ resulting from the Boolean compositions.
When setting $\epsilon_1 = 0$, the strictly active index sets $\mathcal{I}^{0}_{s}$ and $\mathcal{I}^{0}_{n}$ are recovered.
The decision variables of \eqref{BNCBF QP} include the control input $u \in \mathcal{U}$ and the derivatives $\dot{{\lambda}}^{\mathscr{a}}_{j}$ and $\dot{{\lambda}}^{\mathscr{b}}_{j}$ of the optimal dual variables ${{\lambda}}^{\mathscr{a}*}_{j}$ and ${{\lambda}}^{\mathscr{b}*}_{j}$ corresponding to constraint function $h_j$, $j \in \mathcal{I}^{\epsilon_1}_{n}$, defined in the form of \eqref{min distance of two polytopes}.
Note that $\lambda^{\mathscr{a}*}_{j}$ and $\lambda^{\mathscr{b}*}_{j}$ are obtained by evaluating $h_j$, $j \in \mathcal{I}^{\epsilon_1}_{n}$, prior to solving \eqref{BNCBF QP}.
The cost function, with $Q \succ 0$ the weight matrix, enforces the nominal input $u_r$ in Problem \ref{problem 1} is minimally modified.

In constraints \eqref{diff cons a} and \eqref{poly cons a}, $\alpha$ is a locally Lipschitz extended class-$\mathcal{K}$ function.
Constraint \eqref{diff cons a} requires the input $u$ to steer NCBF $h_i$ in a safe direction.
With $\nabla h_i$ known from Assumption \ref{diff asumption}, this condition becomes analogous to the sufficient condition in \eqref{prop 2 condition} for smooth CBFs.
Constraints \eqref{poly cons a}-\eqref{poly cons c} are inspired by \cite{dual_full} and originate from the LP problem \eqref{dual collision avoidance derivative problem}, where \eqref{poly cons a}, \eqref{poly cons b}, and \eqref{poly cons b} correspond to \eqref{LP a}, \eqref{LP a}, and \eqref{LP c}, respectively. 
Together, \eqref{poly cons a}-\eqref{poly cons c} enforce $\dot{L}_j (u, \dot{{\lambda}}^{\mathscr{a}}_{j}, \dot{{\lambda}}^{\mathscr{b}}_{j} ) \geq -\alpha({h}_g(x))$ to be a lower-bound of $\dot{h}_j(x)$ so as to guarantee $\dot{h}_j(x) \geq -\alpha({h}_g(x))$.
Differing from $K^{0}(x)$ in \eqref{LP c}, the index set of almost-active hyperplanes $K^{\epsilon_2}_j(x)$ in \eqref{poly cons b}, defined as
\begin{align}
    K^{\epsilon_2}_j(x) := \{ (k^{\mathscr{a}}, k^{\mathscr{b}}) \mid {\lambda}^{\mathscr{a}*}_{j,\{k^{\mathscr{a}}\}} \leq \epsilon_2, {\lambda}^{\mathscr{b}*}_{j,\{k^{\mathscr{b}}\}} \leq \epsilon_2\},
\end{align}
with $\epsilon_2 > 0$ a small constant, accounts for the nonsmoothness in $h_j$ resulting from the switching of active hyperplanes.

At every time $t \geq 0$, the component NCBFs $h_i$, $i \in \mathcal{I}_s$ and $h_j$, $j \in \mathcal{I}_n$ are evaluated, the corresponding dual variables $\lambda^{\mathscr{a}*}_{j}$ and $\lambda^{\mathscr{b}*}_{j}$ are obtained, and the index sets $\mathcal{I}^{\epsilon_1}_{s}(x(t))$, $\mathcal{I}^{\epsilon_1}_{n}(x(t))$, and $K^{\epsilon_2}_j(x(t))$ are evaluated.
Then, \eqref{BNCBF QP} is solved to obtain $u(x(t)):= u^*(x(t))$.
%
%
%
The following theorem shows $h_g$ is a valid NCBF for \eqref{global dynamics}.

\begin{theorem}\label{main theorem}
Let Assumption 1 hold for all agent geometries ${\mathcal{P}}^i(x^i), i \in \mathcal{N}$.
Let Assumption 2-5 hold.
Define $h_g(x)$ as the composition in \eqref{final bool definition} and suppose there exists $\epsilon_1 > 0$, $\epsilon_2 > 0$, and a locally Lipschitz extended class-$\mathcal{K}$ function $\alpha$ such that the problem in \eqref{BNCBF QP} is feasible for all $x \in \mathcal{C}_g$.
If $x(0) \in \mathcal{C}_g$, then the system \eqref{global dynamics} controlled by $u^*(x(t))$ obtained from \eqref{BNCBF QP} satisfies $x(t) \in \mathcal{C}_g$ for all $t \geq 0$. 
 
\begin{proof} 

By Definition \ref{candidate NCBF definition}, Assumption \ref{D and U assumptions} guarantees $h_g$ and its component functions are proper candidate NCBFs

To show $h_g$ as a valid NCBF, {Proposition \ref{bool main proposition}} provides the sufficient condition
\begin{align}   \label{theo 1 main condition}
    \langle z, v\rangle \geq-{\alpha}\left(h\left( x \right)\right), z \in \Phi_h\left(x\right), \; v \in \Phi_f(x).
\end{align}
Using De Morgan's law, $h_g$ can be written as 
  \begin{align}
      h_g = \max(h_i, h_j),  \; i \in \mathcal{I}_{s}, \; j \in \mathcal{I}_n.
  \end{align}
From Proposition 1 in \cite{magnus_lcs} and Assumption \ref{diff asumption}, $h_g$ satisfies 
\begin{align}
    \partial h_g(x) \subset \operatorname{co} \Phi_h\left(x \right),
\end{align}
where 
\begin{align} \Phi_h\left(x \right) :=  \{\nabla h_i\left(x\right), \partial h_j\left(x\right) \mid i \in \mathcal{I}^0_{s}\left(x\right),  j \in \mathcal{I}^0_{n}\left(x\right) \}. 
\end{align}
From \eqref{filipov} and \eqref{diff inclu definition}, it holds that $F(x) \subset \operatorname{co} \Phi_f(x)$, where
\begin{align}
    \Phi_f(x) := L[f+g u].
\end{align}
Given $\Phi_h\left(x\right)$ and $\Phi_f(x)$, the condition \eqref{theo 1 main condition} can be divided into two groups conditions:
     \begin{align}
  &  \left\langle\nabla h_i\left(x\right),  L[f + g u]\right\rangle   \geq -\alpha(h_g(x)), \; i \in \mathcal{I}^0_{s}(x), \label{dif conditions} \\
    &  \left\langle\partial h_j\left(x\right),  L[f + g u]\right\rangle   \geq -\alpha(h_g(x)),  \; j \in \mathcal{I}^0_{n}(x). \label{nondif condition}
\end{align}

We start by showing \eqref{dif conditions} holds. 
The proof for \eqref{dif conditions} mainly follows the proof of Theorem 3 in \cite{magnus_ccta}, which we reproduce for completeness.
From Lemma 1 of \cite{magnus_ccta}, we know there exists $\delta > 0$ such that, for all $x' \in B(x, \delta)$,
\begin{align} 
&  \mathcal{I}^0_{s}\left(x\right) \subset \mathcal{I}^{\epsilon_1}_{s}\left(x'\right), \; \mathcal{I}^0_{n}\left(x \right) \subset \mathcal{I}^{\epsilon_1}_{n}\left(x'\right).
\end{align}
Then, there exists an index $P$ such that the sequence $x_p \rightarrow x$ satisfies $\| x_p - x \| \leq \delta$ for all $p \geq P$.
Using this sequence, for $i \in \mathcal{I}^0_{s}(x)$, we can show:
\begin{align}
& \left\langle\nabla h_i\left(x\right), L[f + g u]\right\rangle+\alpha\left(h_g\left(x\right)\right)=  \lim_{p \rightarrow \infty} \alpha\left(h_g\left(x_p\right)\right)  \nonumber \allowdisplaybreaks   \\
& + \langle\lim _{p \rightarrow \infty} \nabla h_i\left(x_p\right), \lim_{p \rightarrow \infty}\left(f\left(x_p\right) +g\left(x_p\right) u^*\left(x_p\right)\right)\rangle  \label{theorm 1 inter a} \allowdisplaybreaks   \\
& =\lim_{p \rightarrow \infty}\left\langle\nabla h_i\left(x_p\right), f\left(x_p\right) +g\left(x_p\right) u^*\left(x_p\right)\right\rangle \nonumber \allowdisplaybreaks  \\
& \quad +\lim_{p \rightarrow \infty} \alpha\left(h_g\left(x_p\right)\right) \allowdisplaybreaks  \\
& =\lim_{p \rightarrow \infty}\left(\left\langle\nabla h_i\left(x_p\right), f\left(x_p\right)+g\left(x_p\right) u^*\left(x_p\right)\right\rangle+\alpha\left(h_g\left(x_p\right)\right) \right) \nonumber \allowdisplaybreaks \\
& \geq 0, \label{theorm 1 inter b}
\end{align}
where the last inequality holds because constraint \eqref{diff cons a} holds for all $h_i$, $i \in \mathcal{I}^{\epsilon_1}_{s}\left(x_p\right)$ and $\mathcal{I}^0_{s}\left(x\right) \subset \mathcal{I}^{\epsilon_1}_{s}\left(x_p\right)$.

Next, we prove \eqref{nondif condition}, where $h_j$, $j \in \mathcal{I}^{o}_n(x)$ correspond to the active constraints originating from the minimum distance function \eqref{min distance of two polytopes}.
Given Assumption \ref{over bound assumption}, Assumption \ref{polytope regularity assumptiosn} holds for all $h_j$, $j \in \mathcal{I}_{n}$. Thus, property (c) in the proof of \cite[Theorem 1]{dual_full} holds, which can be equivalently expressed as the existence of $\Bar{\delta} > 0$ such that, for all $x' \in B(x, \Bar{\delta})$,
\begin{align} 
  K^{0}_j(x) \subseteq  K^{\epsilon_2}_j(x').
\end{align}
Givens the balls $B(x, \delta)$ and $B(x, \Bar{\delta})$, we know there exists a sub-sequence of $x_p \rightarrow x$, $p \geq \bar{P} \geq P$, such that 
\begin{align}
   K^{0}_j(x) 
   \subseteq  K^{\epsilon_2}_j(x_p), \;\;  \mathcal{I}^0_{n}\left(x \right) \in \mathcal{I}^{\epsilon_1}_{n}\left(x_p\right).
\end{align}

We now consider the sequence $x_p \rightarrow x$, $p \geq \bar{P}$.
Since Assumption 1 holds, $h_j$ is locally Lipschitz continuous (cf. Lemma 4, \cite{dual_tac}).
Then, from Definition \ref{generalized gradient definition}, we know $h_j$ satisfies 
\begin{align}
 \partial h_j(x) \subset \operatorname{co}\{\lim_{p \rightarrow \infty} \nabla h_j\left(x_p\right) \mid x_p \rightarrow x, x_p \notin  S  \cup \bar{S}_{h_j} \}.
\end{align}
Furthermore, from \eqref{filipov} in Definition 1, we have 
\begin{align}
& L[f + g u] \subseteq \operatorname{co} L[f + g u] \\
& = \operatorname{co}\{\lim_{p \rightarrow \infty} f\left(x_p\right)+g\left(x_p\right) u\left(x_p\right): x_p \rightarrow x, x \notin \bar{S}_{F} \cup S \}. \nonumber
\end{align}
With the above convex hulls of $\partial h_j(x)$ and $L[f + g u](x)$, we can apply Lemma 3 in \cite{magnus_lcs} to prove \eqref{nondif condition} holds by showing
\begin{align} \label{inter 1}
   & \langle   \lim_{p \rightarrow \infty} \nabla h_j\left(x_p\right),   \lim_{p \rightarrow \infty} f\left(x_p\right)+g\left(x_p\right) u^*\left(x_p\right) \rangle   \geq     -\alpha(h_g(x)) \nonumber
\end{align}
hold for $j \in \mathcal{I}^0_{n}(x)$.
The left-hand side of the above satisfies
\begin{align}
   &  \langle  \lim_{p \rightarrow \infty} \nabla h_j\left(x_p\right),    \lim_{p \rightarrow \infty}  f\left(x_p\right)+g\left(x_p\right) u^*\left(x_p\right) \rangle \allowdisplaybreaks \\
= & \lim_{p \rightarrow \infty}  \left\langle  \nabla h_j\left(x_p\right),    f\left(x_p\right)+g\left(x_p\right) u^*\left(x_p\right) \right\rangle \allowdisplaybreaks \\
= & \lim_{p \rightarrow \infty}  \dot{h}_j(x_p) \label{mid mid}
\end{align}
for $j \in \mathcal{I}^{\epsilon_2}_{n}(x_p)$ and, thus, for $j \in \mathcal{I}^{0}_{n}(x)$.

At $x_p$, constraint \eqref{LP a} and \eqref{LP b } hold for $j \in \mathcal{I}^{\epsilon_2}_{n}(x_p)$, which means they hold for $j \in \mathcal{I}^{0}_{n}(x)$.
Similarly, constraint \eqref{LP c} holds for $(k^{\mathscr{a}}, k^{\mathscr{b}}) \in  K^{\epsilon_2}(x_p)$ and $j \in \mathcal{I}^{\epsilon_2}_{n}(x_p)$, which means it holds for $(k^{\mathscr{a}}, k^{\mathscr{b}}) \in  K^{0}(x)$ and $j \in \mathcal{I}^{0}_{n}(x)$.
As a result, the bound \eqref{lowerbound equation} holds for $j \in K^{0}(x)$, i.e., 
\begin{align}
    \dot{h}_j(x_p) \geq g_j(x_p,u^*(x_p)).
\end{align}
Using the above inequality, we can lower-bound \eqref{mid mid} to get
\begin{align}
&  \lim_{p \rightarrow \infty}  \dot{h}_j(x_p)  \geq  \lim_{p \rightarrow \infty} g_j(x_p, u^*(x_p)) \label{t1 eq 1} \allowdisplaybreaks \\
= & \lim_{p \rightarrow \infty} \dot{L}_j(u^*(x_p), \dot{\lambda}^{\mathscr{a}*}(x_p), 
\dot{\lambda}^{\mathscr{b}*}(x_p)) \label{t1 eq 2} \allowdisplaybreaks  \\
\geq & \lim_{p \rightarrow \infty}  -\alpha(h_g(x_p)) \label{t1 eq 3} \allowdisplaybreaks  \\
= & -\alpha(h_g(x)), \label{t1 eq 4}
\end{align}
where \eqref{t1 eq 2} holds because of \eqref{LP a} and \eqref{t1 eq 3} holds because constraints \eqref{poly cons a} holds for $j  \in \mathcal{I}^{0}_{n}\left(x\right) \subset \mathcal{I}^{\epsilon_1}_{n}\left(x_p\right)$.
Since \eqref{dif conditions} holds from \eqref{theorm 1 inter b} and \eqref{nondif condition} holds from \eqref{t1 eq 4}, the sufficient condition \eqref{theo 1 main condition} holds and $h_g(x)$ is a valid NCBF for the system \eqref{global dynamics} controlled by $u^*(x(t))$, i.e, $x(0) \in \mathcal{C}_g \Rightarrow x(t) \in \mathcal{C}_g$. 
\end{proof}
\end{theorem}
 
\begin{remark}
    Theorem \ref{main theorem} guarantees if a system starts in a safe set, then it remains in the safe set for all time.
    Larger $\epsilon_1$ and $\epsilon_2$ increase the number of constraints that are considered throughout $\mathcal{D}$, making the conditions of the theorem harder to satisfy, but in practice may also provide greater robustness to the presence of unmodelled disturbances.
\end{remark}


Next, we demonstrate the proposed method on marine MASs with complex relative-pose constraints through simulation and experiment.

\section{Simulation Results} \label{simulation section}

\begin{figure}[t!] 
 \centering
\includegraphics[width=1\hsize]{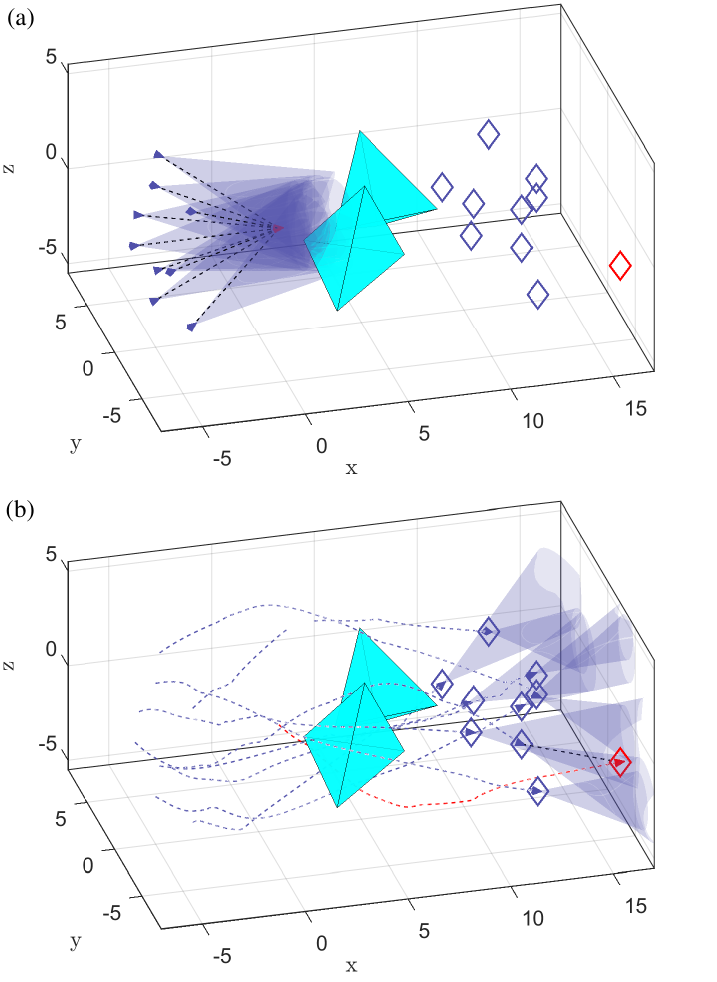}
\caption{Simulated MAS: diamonds represent goal points; the blue and red tetrahedrons represent the follower and leader agents, respectively; the cyan tetrahedrons represent the obstacles; the dotted lines correspond to active LOS connections to the leader; the dotted curves in (b) are trajectories of the agents; (a) $t = 0$ secs and (b) $t = 20$ secs.}
\label{sim_traj}
\end{figure}

In this section, we conduct a simulation on a marine MAS to validate our proposed method. 
We detail the implementation of the method in Section \ref{case study implementation}, whose contents also apply to the experiment in Section \ref{exp section}.
As shown in Fig. \ref{sim_traj}(a), we consider a marine MAS consisting of a leader, nine followers, and two obstacles. Let $\mathcal{N}_F = \{2,\cdots,10\}$ and $\mathcal{N}_O = \{11, 12\}$ be the sets of follower and obstacle indices.

Each follower is equipped with a forward-looking sensor whose origin collocates with the center of the agent. The sensor has a range of $[0.5, 8]$ m, with its FOV modeled as an ellipsoidal cone with half-angle $\phi = 15$°.
We would like to guarantee at least one follower is tracking the leader with its sensor at all times, while all agents navigate toward their goal positions.
 
Assuming no disturbance, the agents share the same dynamics described by the maneuvering model \cite{Fossen2011}:
\begin{subequations} \label{experiment model}
    \begin{align}
  &  \dot{\eta}^i = J^i(\eta^i) \nu^i, \label{kinematics} \\
  &  \mathbf{M}^i \dot{{\nu}}^i+\mathbf{C}^i({\nu}^i) {\nu}^i + \mathbf{D}^i({\nu}^i) {\nu}^i + {g}^i({\eta}^i)+{g}^i_0  = {\tau}^i.  \label{kinetics}
\end{align}
\end{subequations}

In the kinematic model \eqref{kinetics}, the state vector $\eta^i = [x^i,y^i,z^i,\theta^i,\psi^i]^\top$ contains the $x$, $y$, $z$ positions, pitch angle, and yaw angle, respectively, the input vector $\nu^i = [u^i, v^i, w^i, q^i, r^i]^\top$ contains the velocities corresponding to $\eta^i$, and
\begin{align}
    J({\eta^i}) & =  \left[\begin{array}{ccccc} 
R(\theta^i, \psi^i) & \mathbf{0}^{3 \times 1} & \mathbf{0}^{3 \times 1} \\
 \mathbf{0}^{1 \times 3} & 1 & 0 \\
 \mathbf{0}^{1 \times 3}  & 0 & 1/ \mathrm{c} \theta^i 
\end{array}\right],   \label{rotation definition} 
\\  R(\theta^i, \psi^i)^\top & = 
\left[\begin{array}{ccccc}
\mathrm{c} \psi^i \mathrm{c} \theta^i & -\mathrm{s} \psi^i  & \mathrm{c} \psi^i  \mathrm{s} \theta^i \\
\mathrm{s} \psi^i \mathrm{c} \theta^i & \mathrm{c} \psi^i   &  \mathrm{s} \theta^i \mathrm{s} \psi^i \\
-\mathrm{s} \theta^i & 0 & \mathrm{c} \theta^i  
\end{array}\right],
\end{align}
with $s\cdot$, $c\cdot$ the abbreviations of $\sin(\cdot)$ and $\cos(\cdot)$, respectively. 
{In \eqref{kinematics} the roll angles of agents are ignored, as the marine crafts we consider have either active or passive roll-balancing.}
For each agent, the state and velocity vector satisfy $\eta^i \in \mathcal{D}^i := \operatorname{int}\{ \eta^i \mid   - 0.3 \pi   < \psi^i < 0.3 \pi  \}$ and $\nu^i \in \mathcal{U}^i := \{ \nu^i \mid  u^i, v^i, w^i, q^i, r^i \in [-0.2, 0.2]\}$, respectively, with $\operatorname{int}$ denoting the interior of a set.

In the kinetics model \eqref{kinetics}, $\mathbf{M}^i$, $\mathbf{C}^i$, and $\mathbf{D}^i$ are the inertia, Coriolis, and damping matrices, respectively. The vector ${g}^i({\eta^i})$ and ${g}^i_0$ denote the generalized gravitational force and static restoring force, respectively. 
The vector ${\tau}^i = [{\tau}^i_u, {\tau}^i_v, {\tau}^i_w, {\tau}^i_q, {\tau}^i_r]^\top$ contains forces and torques corresponding to the elements in $\nu^i$.
For the simulation, we perform control using the kinematic model \eqref{kinematics} by treating $\eta^i$ as the state and $\nu^i$ as the input, while assuming low-level controllers can generate $\tau^i$ to modify $\nu^i$ sufficiently fast.

The geometries of all agents are described by tetrahedrons ${\mathcal{P}}^i(\eta^i) := \{ p  \mid A^i(\eta^i)  p \leq b^i(\eta^i) \}$ defined as 
\begin{align} 
  & A^i(\eta^i) = \left[\begin{array}{ccccc}
0.24 & 0.84 & 0.48 \\
0.24 & -0.84 & 0.48 \\
-0.97 & 0.00 & 0.00 \\
0.24 & 0.00 & -0.97
\end{array}\right] R(\theta^i, \psi^i)^\top \nonumber \\
& b^i(\eta^i) = [0.06, 0.06,0.24, 0.06]^\top + A^i(\eta^i) p^i. \label{sim agent geometries}
\end{align}

The nominal controller used for each agent $i$ is a proportional controller with inverse kinematics that generates a nominal input $\nu^i_r$ to steer it to its goal state $\eta^i_g$:
\begin{align} \label{ref_controller}
\nu^i_r =  \left[\begin{array}{ccccc}
R(\theta^i, \psi^i) & \mathbf{0}^{3 \times 1} & \mathbf{0}^{3 \times 1} \\
 \mathbf{0}^{1 \times 3} & 1 & 0 \\
 \mathbf{0}^{1 \times 3}  & 0 & 1/ \mathrm{c} \theta^i
\end{array}\right] (\eta^i_g - \eta^i).
\end{align}

\subsection{Encoding of Constraint Functions}\label{case study implementation}
 
In the task considered, for a follower to track the leader, the FOV, LOS, and range constraints should be satisfied. Additionally, collision avoidance constraints, state constraints $\eta^i \in \mathcal{D}^i$, and some regularity-guaranteeing constraints should be enforced.
We encode these constraints as candidate NCBFs.  
 
\vspace{6pt}

\noindent \textbf{Field-of-View Constraints:} To enforce the FOV constraints, we follow the definition in \eqref{FOV constraint definition} and set
\begin{align} \label{ep cone fov def}
   & A^i_{\textup{fov}} = [\mathbf{0}^{2}, I^{2 \times 2}],  c^i_{\textup{fov}}  =[\tan(\phi), 0, 0]^\top, \nonumber  \\
  &  b^i_{\textup{fov}} = d^i_{\textup{fov}} = 0,  p^{ij} = {R}\left( \theta^i,\psi^i \right) (p^j - p^i),
\end{align} 
with ${R}\left( \theta^i,\psi^i \right)$ defined in \eqref{rotation definition}.

 \begin{figure}[t!] 
 \centering
\includegraphics[width=0.85\hsize]{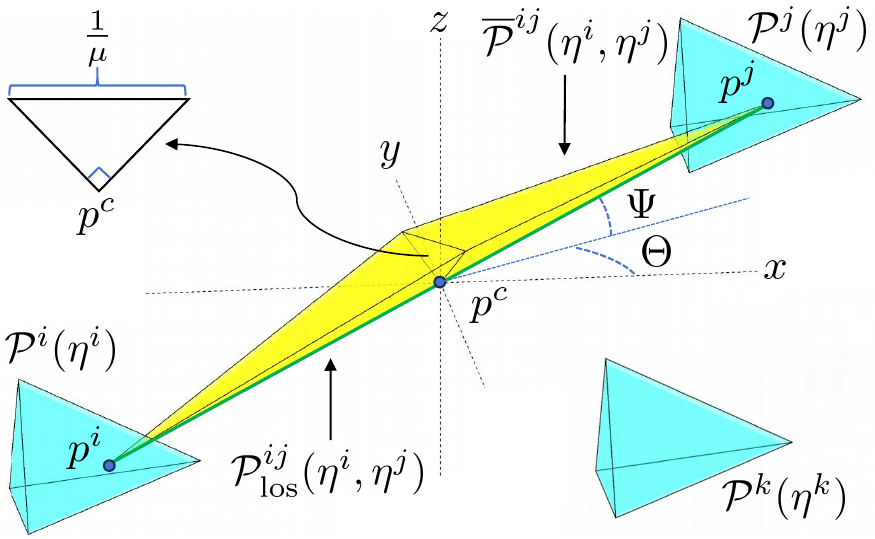}
\caption{Demonstration of the LOS constraint.}
\label{los dem}
\end{figure}

\vspace{6pt}

\noindent \textbf{Line-of-Sight Constraints:} To enforce the LOS constraint \eqref{LOS constraint definition} as $\overline{h}^{ijk}_{\textup{los}}  (x^i,x^j,x^k)$ in \eqref{new LOS min dist definition}, we propose a candidate $\overline{\mathcal{P}}^{ij} (x^i,x^j)$:
\begin{align} \label{LOS poly definition}
    \overline{\mathcal{P}}^{ij}(\eta^i, \eta^j) := \{ p \in \mathbb{R}^3 \mid \overline{A}^{ij}(\eta^i, \eta^j) p \leq \overline{b}^{ij}(\eta^i, \eta^j)\},
\end{align}
where $\overline{b}^{ij}(\eta^i, \eta^j) = \frac{1}{2}\overline{A}^{ij}(\eta^i, \eta^j) ({p^i + p^j}) +  [\frac{1}{2}, \frac{1}{2}, 0, 0]^\top \| p^i - p^j \|$, and 
\begin{align}
  &  \overline{A}^{ij}(\eta^i, \eta^j) = \nonumber \\
  &  \left[\begin{array}{ccccc}
-1 & 0 & \mu \| p^i - p^j \| \\
1 & 0 & \mu \| p^i - p^j \|\\
0 & -1 & -1 \\
0 & 1 & -1 \\
\end{array}\right]  R(  \Theta(\eta^i, \eta^j)  , \Psi(\eta^i, \eta^j))^{\top}, 
\end{align}
with $\Theta(\eta^i, \eta^j) = \operatorname{atan2}({y^i-y^j}, {x^i - x^j})$, $\Psi(\eta^i, \eta^j) = -\operatorname{atan2}({z^i-z^j}, {\sqrt{(x^i - x^j)^2 + (y^i-y^j)^2}})$, and $\mu$ a large positive constant.
The function $\operatorname{atan2}: \mathbb{R} \times \mathbb{R} \rightarrow (-\pi,\pi]$, defined in \cite{MMAS_app_5}, returns the polar angle of the 2-D point $[x,y]^\top$.
Fig. \ref{los dem} demonstrates how $\overline{h}^{ijk}_{\textup{los}}  (x^i,x^j,x^k)$ is implemented, where the cyan tetrahedrons correspond to agent geometries $\mathcal{P}^i(\eta^i)$, $\mathcal{P}^j(\eta^j)$, and $\mathcal{P}^k(\eta^k)$.
The green line segment connecting $p^i$ and $p^j$ represents the LOS set $\mathcal{P}^{ij}(\eta^i,\eta^j)$ and the yellow tetrahedron corresponds to the set $\overline{\mathcal{P}}^{ij} (x^i,x^j)$.
By design, the $(p^i,p^j)$-edge of $\overline{\mathcal{P}}^{ij} (\eta^i,\eta^j)$ always coincides with the LOS set $\mathcal{P}^{ij}_{\textup{los}}(\eta^i, \eta^j)$, implying $\mathcal{C}^{ij}_{\textup{los}}(\eta^i, \eta^j) \subseteq \overline{\mathcal{P}}^{\mathscr{ij}}(\eta^{i},\eta^{j})$.
The cross-section of $\overline{\mathcal{P}}^{ij} (\eta^i,\eta^j)$ is an isosceles right triangle with constant hypotenuse $1/\mu$.
We choose $\mu = 100$ so the tetrahedron is ``slim" and closely approximates $\mathcal{P}^{ij}_{\textup{los}}(\eta^i, \eta^j)$.

\vspace{6pt}

\noindent  \textbf{Range Constraints:} We enforce the range constraints by treating $\underline{h}^{ji}_{\textup{rng}}$ in $\overline{h}^{ji}_{\textup{rng}}$ in \eqref{min range constraint definition} and \eqref{max range constraint definition} as candidate CBFs, with $\underline{r}^{j}_\textup{rng} = 0.5$ and $\overline{r}^{j}_{\textup{rng}} = 8$.

\vspace{6pt}

\noindent  \textbf{Collision Avoidance Constraints:} To enforce the collision avoidance constraints, we choose the minimum distance function $h^{ij }_{\textup{ca}} (x^i,x^j )$ as a candidate NCBF, with safe distance $r^{ij}_{\textup{ca}} = 0.3$ m.

\vspace{6pt}

\noindent \textbf{State Constraints:} We encode $\eta^i \in \mathcal{D}^i$ as 
\begin{align} \label{state cbf}
    h^i_{\mathcal{D}} :=  (0.3 \pi)^2 - { \psi^2 },
\end{align}
and require $h^i_{\mathcal{D}} \geq 0$.

\vspace{6pt}

\noindent \textbf{Regularity constraints:} $\overline{\mathcal{P}}^{ij} (\eta^i,\eta^j)$ in \eqref{LOS poly definition} is well-defined if $(x^i - x^j)^2 + (y^i - y^j)^2 > 0$. We enforce this by introducing
\begin{align}  \label{reg cbf}
    h^{ij}_{\textup{reg}} := (x^i - x^j)^2 + (y^i - y^j)^2 - 0.001
\end{align}
and require $h^{ij}_{\textup{reg}} \geq 0$.

\vspace{6pt}
 
The above NCBFs are then composed as BNCBFs that subsequently form a single BNCBF $h_g$ encoding the global task requirements of the simulation.
Notice Theorem 1 enables the composition of NCBFs using all Boolean operators in \eqref{bool op 1}-\eqref{bool op 3}, as any of their combinations can be converted to a series of $\wedge$ operations through De Morgan's law.
The following notation are useful for constructing compositions: let $\wedge_{i \in \mathcal{N}} h_{i}$ and $\vee_{i \in \mathcal{N}} h_{i}$ denote the $\wedge$ and $\vee$ composition of all functions whose index belongs to $\mathcal{N}$, respectively.
Using composition, the global collision avoidance, state, and regularity constraints are encoded as
\begin{align}
h_{\textup{ca}}  & =  \wedge_{i \in \{1\} \cup \mathcal{N}_F} \left( \wedge_{j \in \{1\} \cup \mathcal{N}_O \cup \mathcal{N}_F \setminus i}  h^{ij}_{\textup{ca}}  \right), \label{collision avoidance comp function definition}  \\
   h_{\mathcal{D}}  &= \wedge_{i \in \{1\} \cup \mathcal{N}_F }  h^i_{\mathcal{D}},  \label{global state cbf} \\
  h_{\textup{reg}} &=  \wedge_{i \in \mathcal{N}_F}   h^{i1}_{\textup{reg}}   ,  \label{global reg cbf}
\end{align}
respectively. The requirement for agent $i \in \mathcal{N}_F$ to track agent $1$ is encoded as
\begin{align}  \label{tra comp function definition}  
    h^{i1}_{\textup{tra}} =  {h}^{i1}_{\textup{fov}}  \wedge\overline{h}^{i1}_{\textup{rng}} \wedge \underline{h}^{i1}_{\textup{rng}} \wedge (\wedge_{k \in \{1\} \cup \mathcal{N}_F  \cup \mathcal{N}_O \setminus \{i,1\}} \overline{h}^{i1k}_{\textup{los}} ),
\end{align}
with $\overline{h}^{i1k}_{\textup{los}}$ specified in \eqref{ep cone fov def}.
With $h_{\mathcal{D}}$, $h_{\textup{reg}}$, $h_{\textup{ca}}$, and $h^{i1}_{\textup{tra}}$, we compose $h_g$ to encode the global task requirements:
\begin{align} \label{hg comp function definition}  
    h_g = h_{\mathcal{D}}  \wedge h_{\textup{reg}} \wedge h_{\textup{ca}} \wedge \left(  \vee_{i \in \mathcal{N}_F} h^{i1}_{\textup{tra}}  \right),
\end{align}
where the last term captures the requirement for at least one follower to track the leader. Let the global safe set be $\mathcal{C}_g := \{ \eta \in \mathcal{D} \mid h_g(\eta) \geq 0\}$, with $\eta$ the global state.

\subsection{Assumption Satisfaction}\label{asp satis}

For Theorem 2 to hold, we must guarantee Assumptions 1-5 hold. 
Assumption \ref{polytope regularity assumptiosn} holds for the agent geometries ${\mathcal{P}}^i(\eta^i)$ because they are tetrahedrons well-defined for all $\eta^i \in \mathcal{D}^i$.
Assumption \ref{polytope regularity assumptiosn} holds for the LOS sets $\overline{\mathcal{P}}^{ij} (\eta^i,\eta^j)$ because they are tetrahedrons well-defined under the constraints in \eqref{reg cbf}. 
Assumption \ref{D and U assumptions} holds by the definition of $D$ and $U$.
Assumption \ref{over bound assumption} holds by construction of $\overline{\mathcal{P}}^{ij} (\eta^i,\eta^j)$, as shown in Fig. \ref{los dem}.
Assumption \ref{non empty assumption} holds by setup, where at least the initial configuration satisfies $\eta(0) \in \mathcal{C}_g$, as shown in Fig. \ref{sim_traj}(a).
$h_g$ can be written in the form of \eqref{final bool definition}, where $h_i$, $i \in \mathcal{I}_{s}$, corresponds \eqref{FOV constraint definition}, \eqref{min range constraint definition}, \eqref{max range constraint definition}, \eqref{state cbf}, and \eqref{reg cbf}.
Within $h_g$, $h_i$, $i \in \mathcal{I}_{s}$ correspond \eqref{FOV constraint definition}, \eqref{min range constraint definition}, \eqref{max range constraint definition}, \eqref{state cbf}, and \eqref{reg cbf}.
Assumption \ref{diff asumption} holds for \eqref{min range constraint definition}, \eqref{max range constraint definition}, \eqref{state cbf}, and \eqref{reg cbf} by construction by construction.
It holds for \eqref{FOV constraint definition} because when \eqref{min range constraint definition} holds the term $\left\|A^i_{\textup{fov}} p^{ij} +b^i_{\textup{fov}}\right\|$ in \eqref{FOV constraint definition} is differentiable.
The feasibility of \eqref{BNCBF QP} is guaranteed since zero input is always a feasible solution for all $\eta \in \mathcal{C}_g$.
Additionally, condition \eqref{ref input condition} holds because the global nominal input $\nu_r = [\nu^{1 \top}_r, \nu^{2 \top}_r, \nu^{3 \top}_r, \nu^{4 \top}_r ,\nu^{5 \top}_r]^\top$ given by \eqref{ref_controller} can steer the system asymptotically to the goal positions.

\subsection{Control Implementation}

Given $\nu_r$, the safe control input $\nu$ is obtained by solving the global BNCBF-QP \eqref{BNCBF QP} originating from $h_g$, with $Q = \textbf{I}^{25}$, $\alpha(s) := 0.2 \cdot s$, and $\epsilon_1 = \epsilon_2 = 0.01$.
The simulation was carried out in MATLAB with a sampling frequency of $10$ hz on a laptop with Intel i7 core and 16 GB RAM. The minimum distance QPs \eqref{min distance of two polytopes} and BNCBF-QP \eqref{BNCBF QP} were solved using MOSEK.

\subsection{Results and Discussions}

\begin{figure}[t!] 
 \centering
\includegraphics[width=1\hsize]{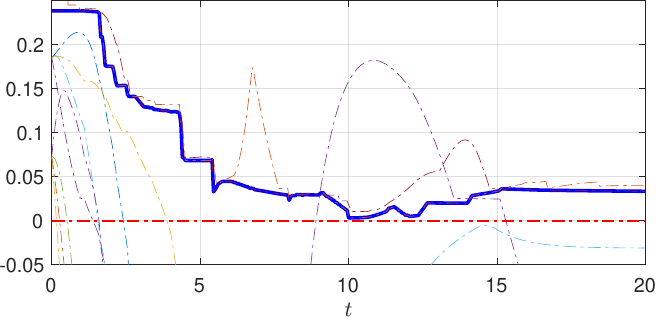}
\caption{Evolution of NCBF: solid blue line and dashed lines represent $h_g$ and the component functions, respectively; dashed red line represents the $0$-level.}
\label{sim_cbf}
\end{figure}

\textit{Constraint Satisfaction and Control Performance:} Fig. \ref{sim_cbf} shows the values of $h_g$ was always above $0$, validating our theoretical guarantees. $h_{\mathcal{D}}$ and $h_{\textup{reg}}$ are omitted as they were inactive.
Fig. \ref{sim_traj} contains snapshots of the system taken at $0$ and $20$ secs, respectively. 
Initially, all followers tracked the leader. 
In the process, the agents took path between the two obstacles due to the need for collision avoidance.
Finally, Fig. \ref{sim_traj}(b) shows the agents converged to their respective goal positions while the leader remained tracked by one follower. 
To conclude, the simulation results verify our theory in Theorem 1, enforcing $h_g \geq 0$ with $u(x(t))$ guaranteed collision avoidance and sensor tracking, while the minimal modification to the nominal controller $u_r(x(t))$ allowed for the stabilization of the system.
 
 \begin{table}[h]
    \centering
    \caption{Problem statistics and solve time (ms) per time step}
    \begin{tabular}{|l|c|c|c|c|}
   \hline $ |\mathcal{N}_{F}| $ & 2 & 5 & 7 & 9\\
   \hline No. of QP \eqref{min distance of two polytopes} & 12 & 42 & 72 & 110 \\
      \hline   $|\mathcal{I}^{\epsilon_1}_{n}| + |\mathcal{I}^{\epsilon_1}_{s}|$   & 1.2 $\pm$ 0.6 & 1.4 $\pm$ 0.7  & 1.7 $\pm$ 1.0 & 2.1 $\pm$ 1.3 \\
\hline Time of all \eqref{min distance of two polytopes} & 11 $\pm$ 0.5 & 38 $\pm$ 0.9 & 64 $\pm$ 1.3 & 98 $\pm$ 1.6 \\
   \hline  Time of \eqref{BNCBF QP}   & 1.1 $\pm$ 0.2 & 1.6 $\pm$ 0.4 & 1.6 $\pm$ 0.5 & 2.2 $\pm$ 0.4 \\ 
   \hline  Total solve time    & 12 $\pm$ 0.7 &  40 $\pm$ 1.3 & 66 $\pm$ 1.8 & 100 $\pm$  2.0 \\ \hline
    \end{tabular}
    \label{comp_time}
\end{table}
 
\textit{Computation Time and Scalability:}

Table \ref{comp_time} compares the computation times between the setup in this simulation (with nine followers) and three other simpler setups to demonstrate the scalability of our method.
The simpler setups are conducted by removing followers from the setup in Fig. \ref{sim_traj}(a).
Each setup is simulated for one trial and the values in rows three to four are averaged between time steps.
As seen, the main computation burden is the total solve time spent on solving the minimum distance QPs \eqref{min distance of two polytopes} originating from the LOS and collision avoidance constraints, whose number increases with complexity $\mathcal{O}( |\mathcal{N}_{F}|^2 + (3 + |\mathcal{N}_{F}|)^2)$.  
On the other hand, the computation time for \eqref{BNCBF QP} is low, as the almost-active set limited the number of constraints (i.e. $|\mathcal{I}^{\epsilon_1}_{n}| + |\mathcal{I}^{\epsilon_1}_{s}|$) included in \eqref{BNCBF QP}.
The last column of Table \ref{comp_time} shows the total solve time 100$\pm$2.0 ms of this simulation.
This shows our method applies to marine MASs with moderate sizes.
For application to larger systems or those with shorter sampling periods, efforts may be made to omit LOS and collision avoidance constraints that are far from being active.



\section{Experiment Results} \label{exp section}

 \begin{figure}[t!] 
 \centering
\includegraphics[width=1\hsize]{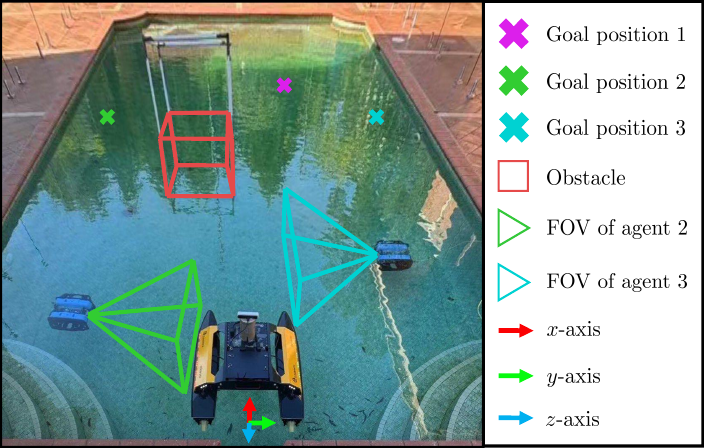}
\caption{Experiment setup}
\label{experiment_env}
\end{figure}

We conduct an experiment on a marine MAS with one USV and two UUVs to highlight the method's practicality. 
The experiment setup is shown in Fig. \ref{experiment_env}, where the marine MAS contains one leader USV and two follower UUVs, and operates in a $12 \times 6 \times 1.2 $ m pool. A $1.2 \times 1.2 \times 1.2$ m cubic obstacle (red box) which we assume blocks camera visions, is located at $[4.6, -0.6, 0.6]$ m. Correspondingly, $\mathcal{N}_F = \{2, 3\}$ and $\mathcal{N}_O = \{4\}$.
Each UUV carries a 1080p camera with a range of $[0.5, 8]$ m, whose FOV can be modeled as a polyhedron cone.
Similar to the simulation, the agents need to reach their respective goal positions while guaranteeing at least one follower is tracking the leader at all times.


The USV has two thrusters, allowing for forward/backward, and yaw motion control. To describe its dynamics with \eqref{experiment model}, we set $z^1, \theta^1, v^1, w^1, q^1, {\tau}^1_v, {\tau}^1_w, {\tau}^1_q = 0$.
The UUVs are fully actuated, keeping zero roll and pitch angles, and operate at a constant depth of 0.5 m. To describe their dynamics with \eqref{experiment model}, we set $z^i = 0.5$, $\theta^i, w^i, q^i, {\tau}^i_w, {\tau}^i_q = 0$, $i \in \mathcal{N}_F$.
The geometries of all agents are described by tetrahedrons ${\mathcal{P}}^i(\eta^i)$ defined in \eqref{sim agent geometries}.
The experiment setup is representative of a range of underwater applications and optical communication, where light signal transmission depends on the LOS, FOV, and range constraints being satisfied \cite{optical_com_survey}.

\subsection{Control Implementation}
 
To construct a global BNCBF $h_g$ capturing the task requirements of the experiment, we follow the definition in \eqref{hg comp function definition} introduced in Section \ref{case study implementation}, with the main difference being the FOV constraints.
To encode FOV of the camera sensors, we follow the definition in \eqref{FOV constraint definition} and define $h^{ij}_{\textup{fov}}$ by setting $A^i_{\textup{fov}} = 0$, $b^i_{\textup{fov}} = 0$, $d^i_{\textup{fov}} = \mathbf{0}^{4}$, and 
\begin{align}  \label{poly cone fov def}
 {c^i_{\textup{fov}}} = 
\left[\begin{array}{rrr}
    0 & -0.64 & -0.77 \\
0.83 & -0.00 & -0.56 \\
-0.83 & 0.00 & -0.56 \\
0 & 0.64 & -0.77
\end{array}\right]^{\top},
\end{align}
where, we slightly abuse notation and define $ {c^i_{\textup{fov}}}$ in a matrix form. Essentially, $h^{ij}_{\textup{fov}}$ capturing a polyhedron cone is the \textup{AND} composition of four constraints representing the hyperplanes of the cone.

\begin{figure}[t!] 
 \centering
\includegraphics[width=0.95\hsize]{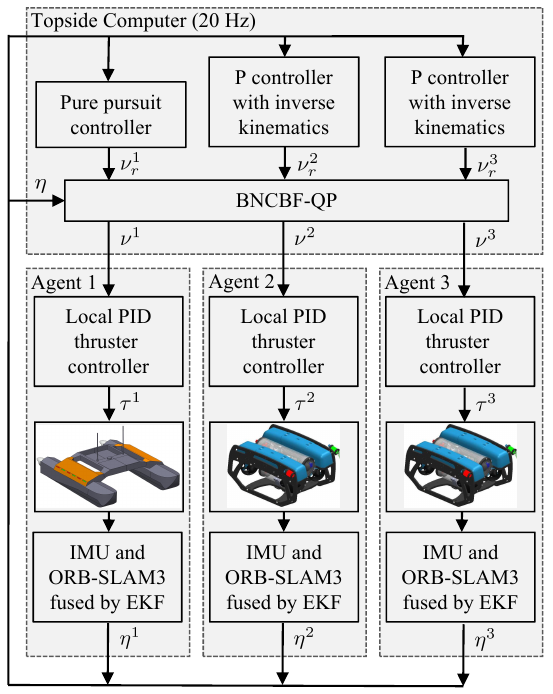}
\caption{Control architecture of the marine MAS}
\label{experiment_archetechture}
\end{figure}  

The control architecture is shown in Fig. \ref{experiment_archetechture}, where the agents are linked to a central computer.
A pure pursuit controller is used for the USV (Agent 1), and the controller defined in \eqref{ref_controller} is used for the UUVs (Agents 2 and 3).
With the state $\eta = [{\eta^1}^\top, {\eta^2}^\top, {\eta^3}^\top]^\top$ and nominal input $\nu_r = [{\nu^1_r}^\top, {\nu^2_r}^\top, {\nu^3_r}^\top]^\top$, we formulate the BNCBF-QP \eqref{BNCBF QP} based on $h_g$ and obtain velocity commands ${\nu^1}$, ${\nu^2}$, and ${\nu^3}$. 
Assumptions 1-5 hold for the same reasons described in Section \ref{asp satis}
The low-level controllers then determine thruster forces ${\nu^1}$, ${\nu^2}$, and ${\nu^3}$ for tracking the reference velocity sufficiently fast.
For state estimation, each agent runs an extended Kalman filter that fuses pose measurements from a visual SLAM algorithm named ORB-SLAM3 \cite{orbslam3} and acceleration measurements from onboard IMUs.
To align the pose measurements of all agents in the same frame, we use a fixed dock with known relative-poses to initialize the agents.
The central computer ran the control programs at 20 Hz, with a Ubuntu 18 system, an Intel i7 core, and 16GB RAM. The minimum distance QPs \eqref{min distance of two polytopes} and BNCBF-QP \eqref{BNCBF QP} were solved using qpsolvers in Python. We did not encounter solve-time issues with this setup.

\subsection{Results and Discussions} 
 
\begin{figure}[t!] 
 \centering
\includegraphics[width=0.95\hsize]{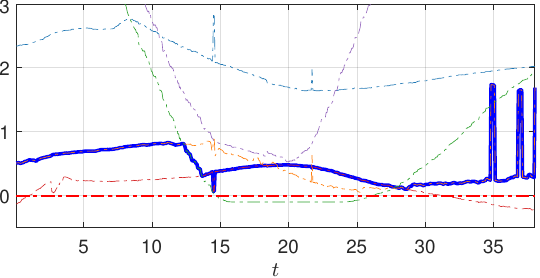}
\caption{Evolution of NCBF: solid blue line and dashed lines represent $h_g$ and the component functions, respectively; dashed red line represents $0$-level.}
\label{exp1_plot}
\end{figure}

\begin{figure}[t!] 
 \centering
\includegraphics[width=1\hsize]{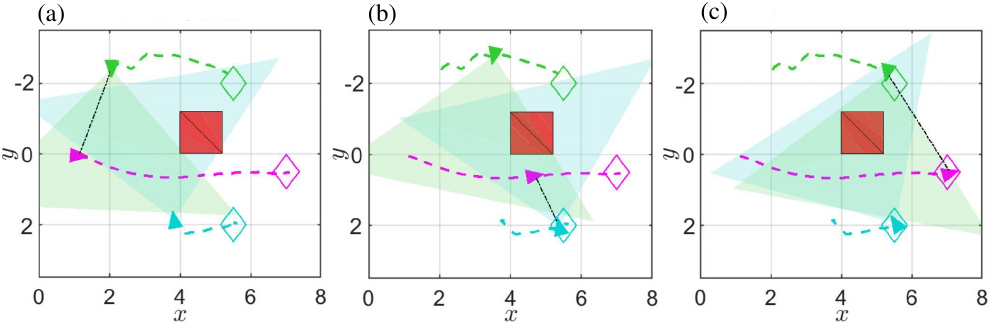}
\caption{Snapshots of the marine MAS: purple, green, and cyan tetrahedrons represent agents 1, 2, and 3, respectively; the red square represents the obstacle; the black line corresponds to active LOS connections to agent 1; dotted curves correspond to the agents' trajectories; diamonds represent goal positions; (a) $t = 0$ secs, (b) $t = 23$ secs, and (c) $t = 38$ secs.}
\label{exp1_traj}
\end{figure}

\textit{Constraint Satisfaction and Control Performance:} Fig. \ref{exp1_plot} shows $h_g > 0$ throughout the course, despite some sudden changes in the function values resulting from state estimation errors, validating the efficacy of our method.
This efficacy can be confirmed by Fig. \ref{exp1_traj}, which plots the bird-eye view trajectories of the marine MAS.
It shows the agents converged to close neighborhoods of their goal positions while at least one UUV tracked the USV with its sensor, where FOV, LOS, and range constraints are guaranteed.

\section{Conclusion}

In this work, a CBF-based framework was proposed to systematically consider complex relative-pose constraints arising in marine MASs, by combing all the constraints as a single NCBF through Boolean composition.
Within the constraints considered, the LOS and collision avoidance constraints were encoded by a dual-based collision avoidance NCBF method. 
Existing safe control design methods are not applicable when such NCBFs are included in the composition.
To address this challenge, we proposed a QP-based safe control design method and developed a new theory to guarantee the resulting controller guarantees the safety of the closed-loop system.
We demonstrated and validated the feasibility and scalability of the approach on marine MASs through both simulation and experiment.
To enhance the method's applicability to large-scale systems or those with short sampling times, future work can be done to implement a distributed version of the proposed method to utilize local computation resources.





\bibliographystyle{IEEEtran}
\bibliography{IEEEabrv,MC}

\begin{thebibliography}{10}
\providecommand{\url}[1]{#1}
\csname url@samestyle\endcsname
\providecommand{\newblock}{\relax}
\providecommand{\bibinfo}[2]{#2}
\providecommand{\BIBentrySTDinterwordspacing}{\spaceskip=0pt\relax}
\providecommand{\BIBentryALTinterwordstretchfactor}{4}
\providecommand{\BIBentryALTinterwordspacing}{\spaceskip=\fontdimen2\font plus
\BIBentryALTinterwordstretchfactor\fontdimen3\font minus \fontdimen4\font\relax}
\providecommand{\BIBforeignlanguage}[2]{{%
\expandafter\ifx\csname l@#1\endcsname\relax
\typeout{** WARNING: IEEEtran.bst: No hyphenation pattern has been}%
\typeout{** loaded for the language `#1'. Using the pattern for}%
\typeout{** the default language instead.}%
\else
\language=\csname l@#1\endcsname
\fi
#2}}
\providecommand{\BIBdecl}{\relax}
\BIBdecl

\bibitem{MMAS_app_survey}
H.~S. Lim, A.~Filisetti, A.~Marouchos, K.~Khosoussi, and N.~Lawrance, ``Applied research directions of autonomous marine systems for environmental monitoring,'' in \emph{OCEANS 2023 - Limerick}, 2023, pp. 1--10.

\bibitem{MMAS_app_5}
D.~S. Terracciano, R.~Costanzi, V.~Manzari, M.~Stifani, and A.~Caiti, ``Passive bearing estimation using a 2-d acoustic vector sensor mounted on a hybrid autonomous underwater vehicle,'' \emph{IEEE Journal of Oceanic Engineering}, vol.~47, no.~3, pp. 799--814, 2022.

\bibitem{MMAS_paper_1}
J.~Lv, Y.~Wang, S.~Wang, X.~Bai, R.~Wang, and M.~Tan, ``A collision-free planning and control framework for a biomimetic underwater vehicle in dynamic environments,'' \emph{IEEE/ASME Transactions on Mechatronics}, vol.~28, no.~3, pp. 1415--1424, 2023.

\bibitem{MMAS_paper_3}
D.~Panagou, S.~Maniatopoulos, and K.~J. Kyriakopoulos, ``Control of an underactuated underwater vehicle in 3d space under field-of-view constraints,'' \emph{IFAC Proceedings Volumes}, vol.~45, pp. 25--30, 2012.

\bibitem{formation_control_1_angle_rigid}
L.~Chen, M.~Cao, and C.~Li, ``Angle rigidity and its usage to stabilize multiagent formations in 2-d,'' \emph{IEEE Transactions on Automatic Control}, vol.~66, no.~8, pp. 3667--3681, 2021.

\bibitem{formation_control_2_dist_rigid}
F.~Mehdifar, C.~P. Bechlioulis, F.~Hashemzadeh, and M.~Baradarannia, ``Prescribed performance distance-based formation control of multi-agent systems,'' \emph{Automatica}, vol. 119, p. 109086, 2020.

\bibitem{opt_1_los_mix}
A.~Caregnato-Neto, M.~R. O.~A. Maximo, and R.~J.~M. Afonso, ``A novel line of sight constraint for mixed-integer programming models with applications to multi-agent motion planning,'' in \emph{2023 European Control Conference (ECC)}, 2023, pp. 1--6.

\bibitem{robot_1_vision}
K.~He, R.~Newbury, T.~Tran, J.~Haviland, B.~Burgess-Limerick, D.~Kulić, P.~Corke, and A.~Cosgun, ``Visibility maximization controller for robotic manipulation,'' \emph{IEEE Robotics and Automation Letters}, vol.~7, no.~3, pp. 8479--8486, 2022.

\bibitem{CBF_survey_ames}
A.~D. Ames, S.~Coogan, M.~Egerstedt, G.~Notomista, K.~Sreenath, and P.~Tabuada, ``Control barrier functions: Theory and applications,'' in \emph{2019 18th European Control Conference (ECC)}, 2019, pp. 3420--3431.

\bibitem{cbf_3_wide_fov}
X.~Li, Y.~Tan, J.~Tang, and X.~Chen, ``Task-driven formation of nonholonomic vehicles with communication constraints,'' \emph{IEEE Transactions on Control Systems Technology}, vol.~31, pp. 442--450, 2023.

\bibitem{cbf_2_los_visual_servoing}
Y.~Zhang, Y.~Yang, and W.~Luo, ``Occlusion-free image based visual servoing using probabilistic control barrier certificates,'' \emph{ArXiv}, vol. abs/2309.03476, 2023.

\bibitem{Nagumo1942berDL}
M.~Nagumo, ``{\"U}ber die lage der integralkurven gew{\"o}hnlicher differentialgleichungen,'' 1942.

\bibitem{Xu2018ConstrainedCO}
X.~Xu, ``Constrained control of input-output linearizable systems using control sharing barrier functions,'' \emph{Autom.}, vol.~87, pp. 195--201, 2018.

\bibitem{ShawCortez2022ARM}
W.~S. Cortez, X.~Tan, and D.~V. Dimarogonas, ``A robust, multiple control barrier function framework for input constrained systems,'' \emph{IEEE Control Systems Letters}, vol.~6, pp. 1742--1747, 2022.

\bibitem{Black2022AdaptationFV}
M.~Black and D.~Panagou, ``Adaptation for validation of a consolidated control barrier function based control synthesis,'' \emph{ArXiv}, vol. abs/2209.08170, 2022.

\bibitem{magnus_lcs}
P.~Glotfelter, J.~Cortés, and M.~Egerstedt, ``Nonsmooth barrier functions with applications to multi-robot systems,'' \emph{IEEE Control Systems Letters}, vol.~1, no.~2, pp. 310--315, 2017.

\bibitem{magnus_ccta}
------, ``Boolean composability of constraints and control synthesis for multi-robot systems via nonsmooth control barrier functions,'' in \emph{2018 IEEE Conference on Control Technology and Applications (CCTA)}, 2018, pp. 897--902.

\bibitem{magnus_tac}
------, ``A nonsmooth approach to controller synthesis for boolean specifications,'' \emph{IEEE Transactions on Automatic Control}, vol.~66, no.~11, pp. 5160--5174, 2021.

\bibitem{jorge_nonsmooth}
J.~Cortes, ``Discontinuous dynamical systems,'' \emph{IEEE Control Systems Magazine}, vol.~28, no.~3, pp. 36--73, 2008.

\bibitem{dual_full}
A.~Thirugnanam, J.~Zeng, and K.~Sreenath, ``Duality-based convex optimization for real-time obstacle avoidance between polytopes with control barrier functions,'' in \emph{2022 American Control Conference (ACC)}, 2022, pp. 2239--2246.

\bibitem{MMAS_survey_2}
Y.~Yang, Y.~Xiao, and T.~shan Li, ``A survey of autonomous underwater vehicle formation: Performance, formation control, and communication capability,'' \emph{IEEE Communications Surveys \& Tutorials}, vol.~23, pp. 815--841, 2021.

\bibitem{Zhang2023AdaptiveNC}
J.~Zhang, X.~Xiang, W.~Li, and Q.~Zhang, ``Adaptive neural control of flight-style auv for subsea cable tracking under electromagnetic localization guidance,'' \emph{IEEE/ASME Transactions on Mechatronics}, vol.~28, pp. 2976--2987, 2023.

\bibitem{Wang2023ModelPT}
K.~Wang, W.~Zou, R.~Ma, Y.~Wang, and H.~Su, ``Model predictive trajectory tracking control of an underactuated bionic underwater vehicle,'' \emph{IEEE/ASME Transactions on Mechatronics}, 2023.

\bibitem{ames_composition_paper}
T.~G. Moln{\'a}r and A.~Ames, ``Composing control barrier functions for complex safety specifications,'' \emph{IEEE Control Systems Letters}, vol.~7, pp. 3615--3620, 2023.

\bibitem{dual_tac}
A.~Thirugnanam, J.~Zeng, and K.~Sreenath, ``Nonsmooth control barrier functions for obstacle avoidance between convex regions,'' 2023.

\bibitem{Fossen2011}
T.~I. Fossen, \emph{{Handbook of Marine Craft Hydrodynamics and Motion Control}}, 2011.

\bibitem{optical_com_survey}
Z.~Zeng, S.~Fu, H.~Zhang, Y.~Dong, and J.~Cheng, ``A survey of underwater optical wireless communications,'' \emph{IEEE Communications Surveys \& Tutorials}, vol.~19, no.~1, pp. 204--238, 2017.

\bibitem{orbslam3}
C.~Campos, R.~Elvira, J.~J.~G. Rodríguez, J.~M. M.~Montiel, and J.~D.~Tardós, ``Orb-slam3: An accurate open-source library for visual, visual–inertial, and multimap slam,'' \emph{IEEE Transactions on Robotics}, vol.~37, no.~6, pp. 1874--1890, 2021.

\end{thebibliography}

\vspace{-33pt}
\begin{IEEEbiography}[{\includegraphics[width=1in,height=1.25in,clip,keepaspectratio]{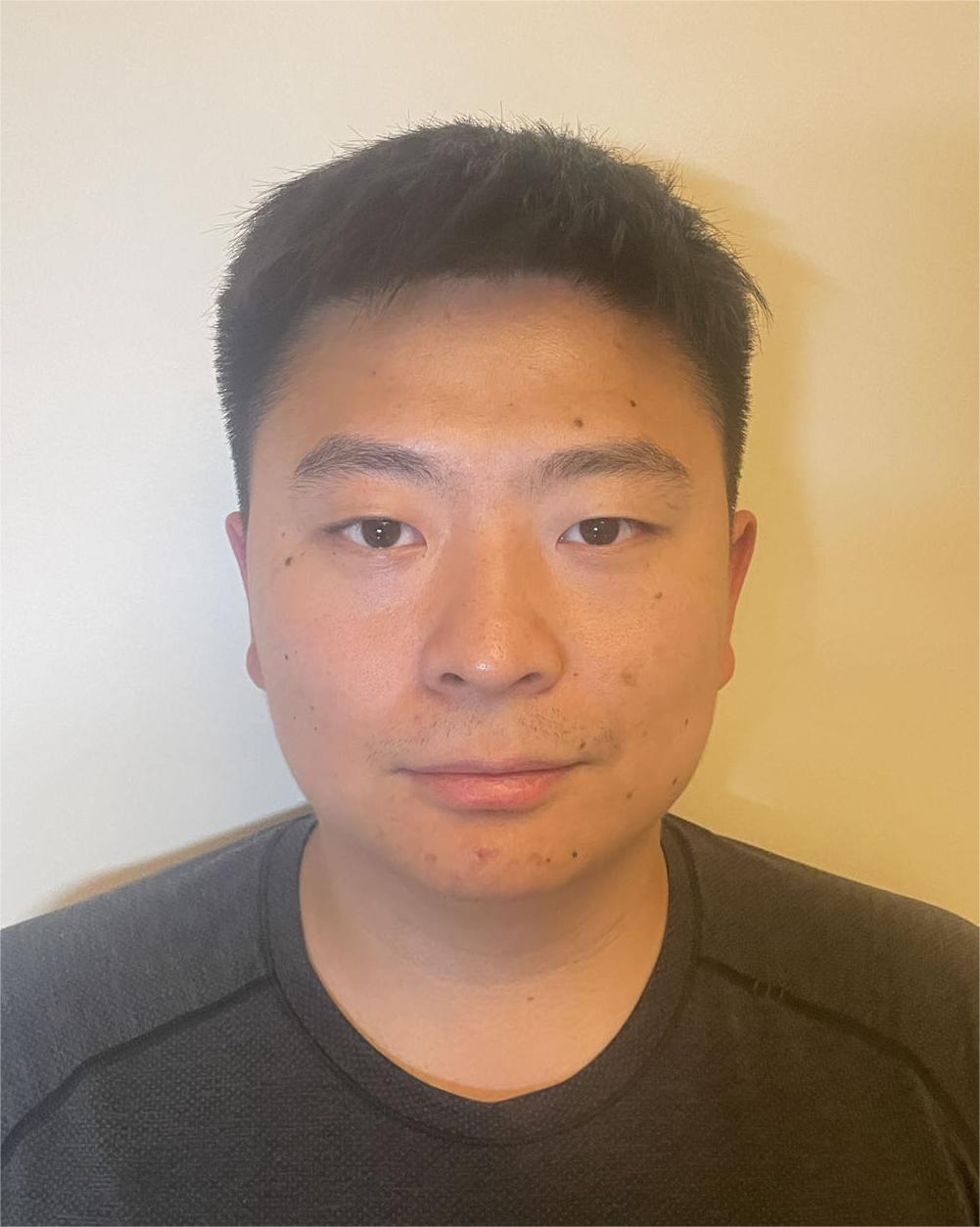}}]{Yujia Yang}
received the B.S. and M.S. degrees in mechanical engineering from Georgia Institute of Technology, Atlanta, USA in 2015 and 2017, respectively. Since 2020, he has been working toward a PhD degree in electrical and electronic engineering at the University of Melbourne, Melbourne, Australia. His research interests include model predictive control, robotics, underwater multi-agent systems, and control barrier function. 
\end{IEEEbiography}

\vspace{11pt}

\vspace{-33pt}
\begin{IEEEbiography}[{\includegraphics[width=1in,height=1.25in,clip,keepaspectratio]{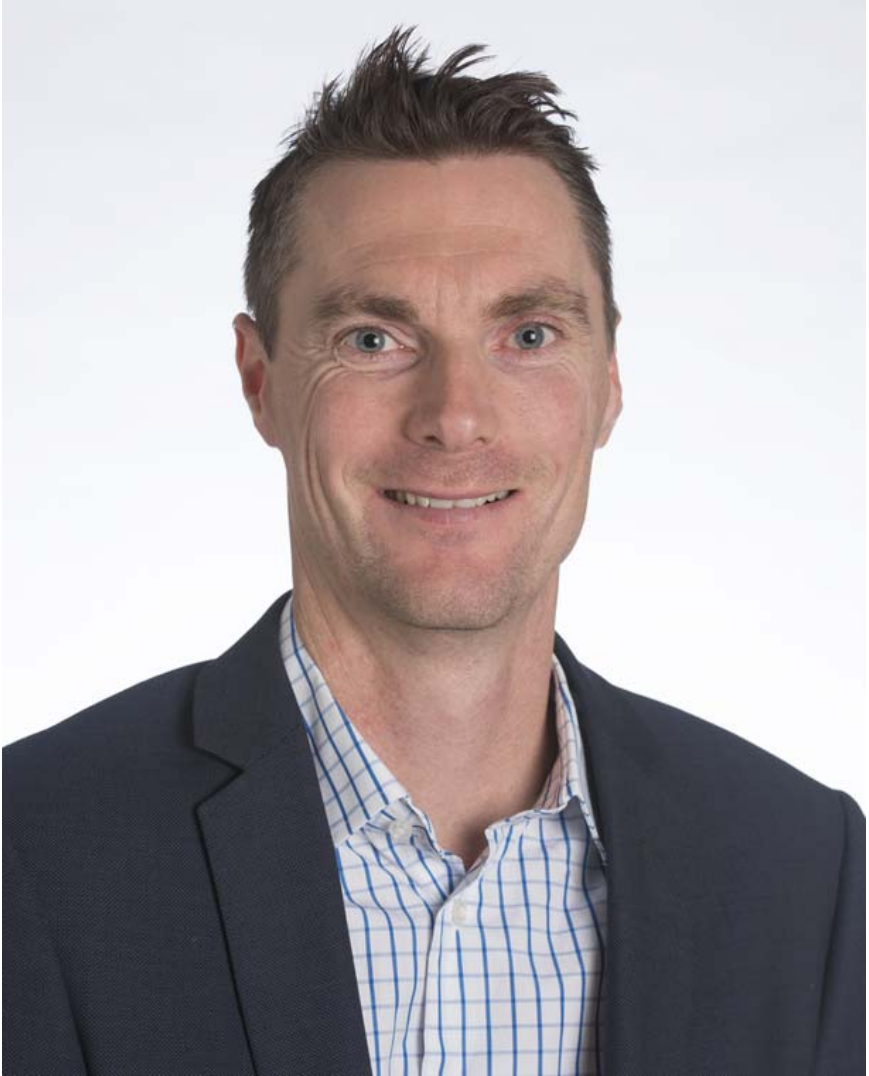}}]{Chris Manzie}
(Senior Member, IEEE,FIEAust) is a Professor and the Head of the Department of Electrical and Electronic Engineering, University of Melbourne, where he is also the Director of the Melbourne Information, Decision, and Autonomous Systems (MIDAS) Laboratory. He was a Visiting Scholar with the University of California at San Diego in 2007 and a Visiteur Scientifique with IFP Energies Nouvelles, Rueil Malmaison, France, in 2012.   His research interests are in model-based and model-free control and optimization with applications in a range of areas, including systems related to autonomous systems, energy, transportation, and mechatronics.
\end{IEEEbiography}

\vspace{11pt}

\vspace{-33pt}
\begin{IEEEbiography}[{\includegraphics[width=1in,height=1.25in,clip,keepaspectratio]{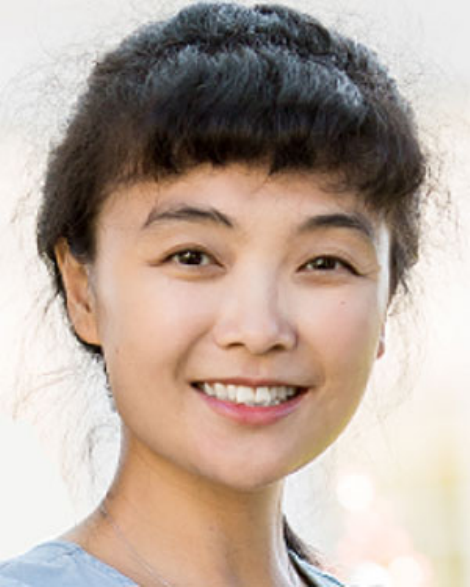}}]{Ye Pu}
received the B.S. degree in electrical engineering from Shanghai Jiao Tong University,
Shanghai, China, in 2008, the M.S. degree in
electrical engineering and computer sciences
from the Technical University Berlin, Berlin, Germany, in 2011, and the Ph.D. degree in electrical
engineering from the Swiss Federal Institute of
Technology Lausanne, Lausanne, Switzerland,
in 2016. She was a Swiss NSF Early Postdoc. Mobility
Fellow and a Postdoctoral Researcher with the
Department of Electrical Engineering and Computer Sciences, University of California, Berkeley, Berkeley, collision avoidance, USA, from 2016 to 2018. She
is currently a Senior Lecturer (Assistant Professor) with the Department of
Electrical and Electronic Engineering, University of Melbourne, Parkville,
VIC, Australia. Her current research interests include learning-based
control, optimization algorithms, and multiagent systems with applications to underwater robotics and energy distribution systems.
\end{IEEEbiography}

\vfill

\end{document}